\documentstyle[prd,eqsecnum,aps,preprint,epsf,tighten,rotate]{revtex}

\begin{document}

\title{A New WIMP Population in the Solar System and New Signals for
Dark-Matter Detectors}
\author{Thibault Damour}
\address{Institut des Hautes Etudes Scientifiques, 91440 
Bures-sur-Yvette, France \\
and DARC, Observatoire de Paris-CNRS, F-92195 Meudon, France}
\author{Lawrence M. Krauss}
\address{Departments of Physics and Astronomy, Case Western Reserve
University\\ 10900 Euclid Ave. Cleveland OH 44106-7079 }

\maketitle

\begin{abstract}
We describe in detail how perturbations due to the planets can cause a
sub-population of WIMPs captured by scattering in surface layers of the Sun to
evolve to have orbits that no longer intersect the Sun.  We argue that such
WIMPs, if their orbit has a semi-major axis less than 1/2 of Jupiter's, can
persist in the solar system for cosmological timescales.  This leads to a new,
previously unanticipated WIMP population intersecting the Earth's orbit.  The
WIMP-nucleon cross sections required for this population to be significant are
precisely those in the range predicted for SUSY dark matter, lying near the
present limits obtained by direct underground dark matter searches using
cyrogenic detectors. Thus, if a WIMP signal is observed in the next generation
of detectors, a potentially measurable signal due to this new population must
exist. This signal, lying in the keV range for Germanium detectors, would be
complementary to that of galactic halo WIMPs.  A comparison of event rates,
anisotropies, and annual modulations would not only yield additional
confirmation that any claimed signal is indeed WIMP-based, but would also allow
one to gain information on the nature of the underlying dark matter model.
\end{abstract}

\section{Introduction}

It is by now firmly established that the dynamics of galaxies and clusters of
galaxies is governed by the existence of large amounts of dark matter, with an
average density sufficient to result in a total mass density in excess of 10$\%$
of the closure density today \cite{dm}.
  At the same time, constraints from Big Bang
Nucleosynthesis \cite{tytler,copist,krausskern} suggest that the abundance
of baryons, including dark baryons, is not likely to be sufficient to account
for all of this material.  Thus, one is led to the possibility of non-baryonic
dark matter, composed of a dissipationless gas of very weakly interacting
elementary particles.  The data from structure formation also suggests that
most of the dark matter must be ``cold", namely non-relativistic at the time
fluctuations on the scale of galaxies first could begin to condense. 

For these reasons, a favored candidate for dark matter is a so-called
WIMP, Weakly Interacting Massive Particle.  The best motivated among the
possibilities involves  supersymmetric extensions of the standard model.  In
these models, the lightest supersymmetric partner of ordinary particles,
usually the neutralino, can be absolutely stable.  Moreover, in order to
resolve various naturalness problems in the Standard Model, the mass scale of
the neutralino is expected to be comparable to the weak symmetry breaking
scale.  As a result, there is a dynamical argument that naturally leads to a
remnant neutralino WIMP abundance comparable to the closure density today.
(e.g. see \cite{jungman}):  The fraction of the closure density
provided by cold relics left over after out-of-equilibrium annihiliation of
an initially thermal distribution of particles, say
$X$, is on the order of
$m_X \, n_X / \rho_{\rm closure} = 
\Omega_X \, \sim 10^{-37} \, {\rm cm}^2 h^{-2}/ \langle \sigma_a (v/c) 
\rangle$ where $\sigma_a$ is the annihilation cross section (and where 
$\Omega_X \equiv \rho_X / \rho_{\rm closure}$, and $h \equiv H_0 / 100 \, 
{\rm km/s} / {\rm Mpc}$) \cite{jungman,kolbturner}.As a result,
the typical annihilation cross section  needed for leaving a density of massive
relics comparable to the closure  density is a {\it weak scale} cross section:
$\alpha^2 (100 \, {\rm  GeV})^{-2} \sim (\alpha / 10^{-2})^2 \times 4 \times
10^{-36} \, {\rm  cm}^2$. In this paper, we shall assume that the dark matter
is indeed  made of such weakly interacting massive particles (WIMPs). Though
most of our work  depends only on the general assumption of WIMPS (with mass
$10 \, {\rm  GeV} < m_X < 1000 \, {\rm GeV}$), we shall give numerical
estimates for  the effects we discuss by sampling the parameter space of the
Minimal  Supersymmetric Standard Model (MSSM) in its various
forms\cite{jungman}.

Several searches for these WIMPs are underway, using either direct 
(energy deposition in laboratory samples) or indirect (e.g. looking for 
products of WIMP annihilation) techniques. Direct searches look for 
recoil events (with associated phenomena such as heat deposition, 
ionisation, etc.) due to the elastic scattering of WIMPs on nuclei. These 
searches present many experimental challenges because: (i) the rate of 
expected signals is small (less than a few events/day/kg), (ii) the 
recoil energies are also small (typically from 1 to 30 keV), and (iii) 
the background rate is comparatively very high.

Any {\it a priori} theoretical information on the expected signals is 
therefore very important for separating true signals from background 
events. In the present paper, we shall describe in detail the derivation and
characteristics of our previously announced result \cite{damkrauss1} that a
heretofore unexplored  aspect of the dynamical history of WIMPs in the solar
system may lead to a new population of WIMPs whose signals lead to a
peculiar signature in the differential energy  spectrum of recoil events, and,
probably, peculiar anisotropy and annual modulation features.  The new signals
discussed here correspond to an excess of recoil events in an energy window on
the keV energy scale. The ratio of the rate of  these events (per day, per kg
and per keV) to the standardly expected one is proportional to an average
scattering cross section of the WIMP, weighted over elements in the Sun. 
 Sampling over SUSY parameter space, we find that
this event rate ratio can be larger than a factor of 2.  Most important
perhaps, we find that it is for cross sections that lie just below current
experimental limits, namely for those WIMPs that are the prime target of the
next generation of underground detectors, that the new signal we describe is
maximal.  Hence, if WIMPs are discovered in the next generation of detectors,
the signal we propose may be one of the best ways of demonstrating its origin,
and probing the characteristics of the WIMP dark matter responsible for it. 
Finally,  another signal  associated to the new dynamical class of
WIMPs that we discuss here might  be a significant increase in the (indirect)
neutrino signal expected from  WIMP annihilations in the Earth.  Indeed, limits
from such searches might possibly further constrain the region of SUSY dark
matter parameters that remain viable.

The outline of this paper is as follows:  First, we derive in section II
 the differential
capture rate by the Sun for WIMPs that might eventually form part of the
population of interest here.  Next (section III),
 we derive in detail the dynamical
equations that govern the possible diffusion of this population into bound
solar system orbits that no longer intersect the Sun.  In the next section,
we combine the results of the two preceding discussions to estimate the
capture rate of long-surviving solar-system bound WIMPs.  Based on this, we can
then estimate (in section V) the present phase-space distribution 
of such WIMPS in the region
of the Earth. We are then able, in section VI, to explore the
possible observable signals from this new population, in terms of the
differential event rate per kg per day per keV of scattering on target nuclei.
In section VII we explore these results in the context of realistic SUSY
WIMPS.  By sampling the allowed SUSY phase space, using accelerator
constraints, and using the ``Neutdriver" code \cite{jungman} to determine
remnant WIMP densities and calculate capture and scattering cross sections on
various targets, we demonstrate that the direct scattering signal from this new
population should in principle be detectable if WIMPs lie near the current
bounds.  Finally, in the concluding section we examine other possible
signatures for this distribution, and discuss future work that would be
useful to perform in light of the results we present here.

\section{Differential capture of WIMPs by the Sun}

Direct searches of WIMPs assume that the recoil events are due to the 
scattering on a nucleus of a WIMP coming directly from a ``standard'' 
galactic halo, with a local mass density (around the solar system) 
$\rho_X \sim 0.3 \, {\rm GeV} / {\rm cm}^3$, and a roughly Maxwellian 
velocity distribution (in the galactic inertial frame) characterized by 
$v_{\rm rms} \sim 270 \, {\rm km/sec}$. In this paper, we shall study a 
particular class of WIMPs that underwent the following dynamical 
history:

(i) coming from the galactic halo, they are scattered by nuclei in the 
outskirts of the Sun into very elliptic, bound orbits with semi-major 
axis $a \sim 1$ astronomical unit (AU);

(ii) being perturbed by gravitational interaction with the planets, they 
diffuse out of the Sun and stay bound in the inner solar system during 
its entire age, $t_S \sim 4.5 \, {\rm Gyr}$;

(iii) finally, they form a class of low velocity WIMPs, with typical 
barycentric velocities $v_X \sim v_{\rm Earth} \sim 30 \, {\rm km/s}$, 
which can either undergo a scattering event with a nucleus in a 
dark-matter detector, or be scattered by a nucleus in the Earth to be 
ultimately accreted to finally annihilate at the Earth's center.

In this Section, we consider the first step of this process: the 
scattering by nuclei in the Sun into an AU-scale bound orbit. The 
essential new feature with respect to previous analyses of capture of 
WIMPs by the Sun \cite{kraussetal,gould87,gould92}
will  be to concentrate on WIMPs that {\it graze the Sun}, and lose just 
enough energy to stay in Earth-crossing orbits. To estimate the number of 
WIMPs susceptible to be sufficiently perturbed by the small gravitational 
interaction of planets, we need first to derive the {\it differential} 
capture rate, per energy and per angular momentum, of WIMPs by the Sun. 
We shall ultimately be interested only in the small fraction of WIMPs 
which have angular momenta in a small range $J_{\rm min} \le J \le J_S$ 
where $J_S$ is the angular momentum for a WIMP exactly grazing the Sun. 
 The result we need is a simple generalization of results 
previously derived in the literature \cite{gould87,gould92}. However,
for the benefit of the general reader we shall present a self-contained
derivation starting essentially from scratch.

Let
\begin{equation}
dn_X = f_{\infty} ({\bf v}_{\infty}) \, d^3 \, {\bf x}_{\infty} \, d^3 \, 
{\bf v}_{\infty} = f_{\rm loc} ({\bf x}_{\rm loc} , {\bf v}_{\rm loc}) \, 
d^3 \, {\bf x}_{\rm loc} \, d^3 \, {\bf v}_{\rm loc} \, , \label{eq2.1}
\end{equation}
be the phase-space distribution of WIMPs, et infinity (in the galactic 
halo), and locally within the solar system. We shall always refer WIMP 
velocities to a frame attached to the Sun. Let ${\bf v}_S$ denote the 
(vectorial) velocity of the Sun with respect to a galactic frame. The 
galactic WIMP velocity would be
 $ {\bf v}^{\prime}_{\rm WIMP} = {\bf v}_{\rm WIMP} + {\bf v}_S$. The 
distribution function at infinity is taken to be Maxwellian,
\begin{equation}
f_{\infty} ({\bf v}_{\infty}) = \frac{n_X}{\pi^{3/2} \, v_o^3} \ \exp \, 
\left( - \frac{({\bf v}_{\infty} + {\bf v}_S)^2}{v_o^2} \right) 
\label{eq2.2}
\end{equation}
where $v_o^2 \equiv \frac{2}{3} \, v_{\rm rms}^{\prime \, 2}
 \equiv \frac{2}{3} \, \langle {\bf v}_{\infty}^{\prime \, 2} \rangle$
 is an rms ``planar'' velocity.  As the 
class of WIMPs studied here does not depend much on any eventual cut-off 
of $f_{\infty}$ beyond some ``evaporation'' velocity, we shall work with 
the simple exponential form (\ref{eq2.2}). In the body of the text we 
shall assume the following standard values for the parameters $\rho_X = 
n_X \, m_X$, $v_o$, and $v_S$:
\begin{equation}
\rho_X^{\rm standard} = n_X \, m_X = 0.3 \, {\rm GeV} \, {\rm cm}^{-3} \, 
, \ v_o^{\rm standard} = 220 \, {\rm kms}^{-1} \, , \ v_S = 220 \, {\rm 
kms}^{-1} \, . \label{eq2.3}
\end{equation}
At the end, we shall comment on the effect of changes around the standard 
values for $\rho_X$ and $v_o$.

Liouville's theorem tells us that $f_{\rm loc} ({\bf x}_{\rm loc} , {\bf 
v}_{\rm loc})$ is constant along the free motion of WIMPs. Therefore, at 
any point along an incoming trajectory
\begin{equation}
f_{\rm loc} \, ({\bf x}_{\rm loc} , {\bf v}_{\rm loc}) = f_{\infty} \, [ 
{\bf v}_{\infty} ({\bf x}_{\rm loc} , {\bf v}_{\rm loc}) ] \, , 
\label{eq2.4}
\end{equation}
where ${\bf v}_{\infty} ({\bf x}_{\rm loc} , {\bf v}_{\rm loc})$ is the 
incoming velocity at infinity of the WIMP observed locally, at position 
${\bf x}_{\rm loc}$, with velocity ${\bf v}_{\rm loc}$, within the solar 
system (before it undergoes any scattering event). In spherically 
symmetric problems, only the angular average $\overline{f}_{\rm loc} \, 
({\bf x}_{\rm loc} , {\bf v}_{\rm loc}) = \overline{f}_{\infty} \, 
[v_{\infty} ({\bf x}_{\rm loc} , {\bf v}_{\rm loc})]$ matters, with
\begin{equation}
\overline{f}_{\infty} \, (v_{\infty}) = \frac{n_X}{4 \, \pi^{3/2}} \, 
\frac{1}{v_o \, v_S \, v_{\infty}} \, \left[ e^{-\left( \frac{v_{\infty} 
- v_S}{v_o} \right)^2} - e^{-\left( \frac{v_{\infty} + v_S}{v_o} 
\right)^2} \right] \, . \label{eq2.5}
\end{equation}
It is standard to write the differential scattering cross-section of the WIMP
$X$  onto the nucleus of atomic number $A$ as
\begin{equation}
d \, \sigma_A = \sigma_A \, F_A^2 (Q) \, \frac{d \, \Omega_{\rm 
cm}}{4\pi} \, , \label{eq2.6}
\end{equation}
where $Q = E_{\rm before} - E_{\rm after}$ is the energy transferred 
during the scattering, $F_A (Q)$ is a form factor, and $d \Omega_{\rm cm} 
= \sin \theta_{\rm cm} \, d \theta_{\rm cm} \, d \varphi_{\rm cm}$ is the 
scattering solid angle element {\it in the center of mass} frame. We 
shall take an exponential form factor
\begin{equation}
F^2 (Q) = \exp \, (-Q/Q_A) \, , \label{eq2.7}
\end{equation}
where the nucleus-dependent quantity $Q_A$ will be discussed below. Note 
that $\sigma_A$ (often denoted $\sigma_A^0$) in Eq.~(\ref{eq2.6}) is by 
definition independent of the scattering angle.

Let us consider a volume element $d^3 {\bf x}$ in the Sun, containing 
$n_A ({\bf x}) \, d^3 {\bf x}$ nuclei of atomic number $A$. The 
differential {\it flux} of WIMPs impinging on this lump of scatterers is 
$d^3 {\bf v}_{\rm loc} \, f_{\rm loc} ({\bf x} , {\bf v}_{\rm loc}) \, 
\vert {\bf v}_{\rm loc} \vert$. Therefore, the corresponding differential 
number per second of scattering events (within the center-of-mass 
scattering solid angle $d \Omega_{\rm cm}$) reads
\begin{equation}
d \dot{N}_A =  d^3 {\bf x} \, n_A ({\bf x}) \, d^3 {\bf v}_{\rm loc} \, 
f_{\rm loc} ({\bf x} , {\bf v}_{\rm loc}) \, \vert {\bf v}_{\rm loc} 
\vert \, \sigma_A \, F_A^2 (Q) \, \frac{d \Omega_{\rm cm}}{4\pi} \, . 
\label{eq2.8}
\end{equation}
We wish to sort out this scattering rate according to the distribution in 
outgoing semi-major axis $a$ and (specific) angular momentum $J$. The 
conserved energy (kinetic plus potential) of the WIMP before the 
scattering event is
\begin{equation}
E_{\rm before} = \frac{1}{2} \, m_X ({\bf v}_{\rm loc}^2 -  v_{\rm 
esc}^2 (r)) = \frac{1}{2} \, m_X \, {\bf v}_{\infty}^2 \, , \label{eq2.9}
\end{equation}
where ${\bf v}_{\rm loc} \equiv {\bf v}_{\rm before}$ denotes the local 
velocity before the collision, and where $v_{\rm esc}^2 (r) \equiv 2 \, 
U(r) \equiv + 2 \int d^3 {\bf x}' \, G_N \rho ({\bf x}') / \vert {\bf x} 
- {\bf x}' \vert$ is the escape velocity at the radius $r$ within the Sun 
($G_N$ denoting the Newtonian gravitational constant). The conserved 
energy after the collision reads
\begin{equation}
E_{\rm after} = \frac{1}{2} \, m_X ({\bf v}_{\rm after}^2 - v_{\rm 
esc}^2 (r)) = - \frac{G_N \, m_X \, M_{\odot}}{2a} \equiv - \frac{1}{2} \, m_X 
\, \alpha \, , \label{eq2.10}
\end{equation}
where $M_{\odot}$ denotes the mass of the Sun, and where $\alpha$ is a 
shorthand for $G_N \, M_{\odot} / a$. By the standard laws of nonrelativistic 
elastic collisions (see, e.g., \cite{landau}), the velocity after the 
collision, and therefore the energy transfer, are linked to the c.m. 
scattering angle $\theta_{\rm cm}$ by
\begin{equation}
{\bf v}_{\rm after}^2 = {\bf v}_{\rm loc}^2 \, \left[ 1 - \frac{1}{2} \, 
\beta_+^A (1 - \cos \theta_{\rm cm})\right] , Q = E_{\rm before} - E_{\rm 
after} = \frac{1}{2} \, m_X \, \beta_+^A \, {\bf v}_{\rm loc}^2 \, 
\frac{1 - \cos \theta_{\rm cm}}{2} \, , \label{eq2.11}
\end{equation}
where we define (following Ref.~\cite{jungman})
\begin{equation}
\beta_{\pm}^A \equiv \frac{4 \, m_X \, m_A}{(m_X \pm m_A)^2} \, . 
\label{eq2.12}
\end{equation}
Note that the maximum value of $\beta_+^A$ is one, which is reached when 
the mass of the WIMP matches the mass of the nucleus: $m_X = m_A$. For a 
given incoming local velocity ${\bf v}_{\rm loc}$, Eqs.~(\ref{eq2.9}), 
(\ref{eq2.10}) give
\begin{equation}
Q = \frac{1}{2} \, m_X (v_{\rm loc}^2 - v_{\rm esc}^2  (r) + \alpha) = 
\frac{1}{2} \, m_X (v_{\infty}^2 + \alpha) \, . \label{eq2.new}
\end{equation}
On the other hand, Eq.~(\ref{eq2.11}) relates the energy transfer $Q$ to 
the c.m. scattering solid angle. Finally, we have
\begin{equation}
\frac{d \Omega_{\rm cm}}{4\pi} = d \left( \frac{1 - \cos \theta_{\rm 
cm}}{2} \right) = \frac{d \, Q}{\frac{1}{2} \, m_X \, \beta_+^A \, v_{\rm 
loc}^2} = \frac{d \, \alpha}{\beta_+^A \, v_{\rm loc}^2} \, , 
\label{eq2.13}
\end{equation}
where we recall that $\alpha \equiv G_N \, M_{\odot} / a$.

Inserting Eq.~(\ref{eq2.13})   in Eq.~(\ref{eq2.8}) 
yields
\begin{equation}
d \dot{N}_A =  d^3 {\bf x} \, n_A ({\bf x}) \, \sigma_A \, \frac{d^3 
{\bf v}_{\rm loc} \, f_{\rm loc} ({\bf x} , {\bf v}_{\rm loc})}{\beta_+^A \, 
v_{\rm loc}} \, F_A^2 (Q) \, d \alpha \, \Theta_{\alpha} \, , 
\label{eq2.15}
\end{equation}
where the last  factor is a step function ($\Theta (x) \equiv 1$ if 
$x \geq 0$, and vanishes if $x < 0$) taking care of the inequality on 
$v_{\rm loc}$ or $v_{\infty}^2 ({\bf x} , {\bf v}_{\rm loc}) = v_{\rm 
loc}^2 - v_{\rm esc}^2 (r)$ entailed by the constraint $(1 
- \cos \theta_{\rm cm}) / 2 \leq 1$ , i.e. $2Q / 
m_X = v_{\infty}^2 + \alpha \leq \beta_+^A \, v_{\rm loc}^2 = \beta_+^A 
(v_{\infty}^2 + v_{\rm esc}^2 (r))$. 
 Using the identity $(1 / \beta_+^A) - (1 / 
\beta_-^A) \equiv 1$, easily checked from the definitions (\ref{eq2.12}), 
this constraint leads to the step function
\begin{equation}
\Theta_{\alpha} \equiv \Theta \, \left[ \beta_-^A \left(v_{\rm esc}^2 (r) 
- \frac{\alpha}{\beta_+^A} \right) - v_{\infty}^2 \right] \, . 
\label{eq2.16}
\end{equation}

The phase-space distribution of the WIMPs at infinity, Eq.~(\ref{eq2.2}),
is anisotropic. If the  overall capture mechanism discussed in this paper
were spherically symmetric (i.e. a capture by a spherical Sun
followed by a spherically symmetric exit mechanism), it would be exact
to replace, for purposes of capture calculations, the (anisotropic)
incoming phase-space distribution,  Eq.~(\ref{eq2.2}), by its (isotropic)
 angular average, Eq.~(\ref{eq2.5}). Actually, the overall capture 
mechanism discussed here is not
spherically symmetric (even in the (excellent) approximation of an exactly
spherical Sun), because, as we shall see in the next section, two angular
parameters ($i$ and $g$), related to the spatial orientations of ${\bf v}_{\rm
after}$ and ${\bf x}$ with respect to the ecliptic plane, modulate the 
efficiency of extraction of the WIMPs captured by the Sun. However, we
expect that, because of the partial randomisation of the incoming directional
quantities by scattering on a spherical Sun, the effects linked to the overall
lack of spherical symmetry, i.e. the correlation effects between the incoming
anisotropy (linked to the direction of ${\bf v}_S$) and the outgoing one
(linked to the ecliptic plane) are small. Therefore, we shall henceforth 
neglect them, for purposes of capture calculations,
 and replace the anisotropic distribution, Eq.~(\ref{eq2.2}),
by its angular average, and, correspondingly, the local distribution
 $f_{\rm loc} ({\bf x} , {\bf v}_{\rm loc})$ 
by its angular average $\overline{f}_{\rm loc}$, which, thanks to 
Liouville's theorem (\ref{eq2.4}), is simply equal to 
$\overline{f}_{\infty} \, [v_{\infty} (r , v_{\rm loc})]$.
We can then make use of the spherical symmetry of  $\overline f_{\rm loc}$ 
and $ n_A ({\bf x}) = n_A(r)$ to simplify the problem
of sorting out the differential rate (\ref{eq2.15}) according to the 
distribution in the outgoing (specific) angular momentum
  $J = \vert {\bf x} \times {\bf v}_{\rm after} \vert$.
(We thank A. Gould for suggesting this simplification).
Let $\theta$ denote the colatitude of ${\bf v}_{\rm after}$
with respect to the radial direction taken  as $z$-axis, i.e. 
$J = r  v_{\rm after} \sin \theta$. Since an isotropic
distribution is uniform in $\cos \theta$, and $J \propto \sin \theta$,
the distribution in $J^2$ is given by the normalized measure
\begin{equation}
d \, J^2 ( 2  J_{\max}^2)^{-1} (1 - J^2 / J_{\max}^2 )^{-1/2} \Theta_J \, ,
\label{eq2.18new}
\end{equation} 
where
\begin{equation}
 J_{\max}(r, \alpha ) \equiv r ( v_{\rm esc}^2(r,\alpha) - \alpha )^{1/2}= r 
v_{\rm after} \; , \, \Theta_J \equiv \Theta (  J_{\max}(r, \alpha ) - J) \, .
\label{eq2.19new}
\end{equation}
Here, the theta function $ \Theta_J$ takes care of the constraint
$ \sin \theta < 1$. This leads to a differential scattering rate of the form

\begin{equation}
d \dot{N}_A =  d^3 {\bf x} \, n_A (r) \, \sigma_A \, \frac{4\pi v_{\rm loc}
d v_{\rm loc} \, \overline{f}_{\rm loc} (r , v_{\rm loc})}
{2 J_{\max}^2 \, \beta_+^A } \, 
 ( 1 - \frac{ J^2}{ J_{\max}^2})^{-1/2}F_A^2 (Q) \,
 \Theta_{\alpha} \, \Theta_J \,  d \alpha \, dJ^2 \, . \label{eq2.20new}
\end{equation}

 Performing the integral over $J$ in Eq.~(\ref{eq2.20new}) over the range 
$J_{\rm min} \leq J \leq J_{\max}$ leads to
\begin{equation}
d \dot{N}_A \vert_{J \geq J_{\min}} = d^3 {\bf x} \, n_A (r) \, 
\sigma_A \left( 1 - \frac{J_{\min}^2}{(J_{\max}(r,\alpha))^2} \right)^{1/2} 
\, \Theta_{J_{\min}} \, \frac{4\pi \, v_{\rm loc} \, d v_{\rm loc} \, 
\overline{f}_{\rm loc}}{\beta_+^A} \, F_A^2 (Q) \, \Theta_{\alpha} \, d 
\alpha \, , \label{eq2.18}
\end{equation}
in which $Q$ should be replaced by its expression in terms of $v_{\rm 
loc}^2$ or $v_{\infty}^2$, Eq.~(\ref{eq2.new}). Here, $\Theta_{J_{\min}} 
= \Theta \, [r \, v_{\rm after} - J_{\min}] = \Theta  \, 
[ J_{\max}(r,\alpha) - J_{\min}]$ gives a lower bound to the 
admissible values of $r = \vert {\bf x } \vert$.  
The integrals over $ d^3 {\bf x} = 4 \pi r^2 dr$ and $ d v_{\rm loc}$
are uncoupled.  Using the exponential form factor (\ref{eq2.7}) we then 
define the function
\begin{eqnarray}
K_A (r,\alpha) & \equiv & \frac{v_o}{n_X} \, \frac{1}{\beta_+^A} \int 
4\pi \, v_{\rm loc} \, d v_{\rm loc} \, \overline{f}_{\rm loc} \, \exp 
\left[ - \frac{m_X}{2 Q_A} \, (v_{\rm loc}^2 - v_{\rm esc}^2 (r) + 
\alpha) \right] \, \Theta_{\alpha} \nonumber \\
& \equiv & \frac{v_o}{n_X} \, \frac{1}{\beta_+^A} \int 4\pi \, v_{\infty} 
\, d v_{\infty} \, \overline{f}_{\infty} (v_{\infty}) \, \exp \left[ - 
\frac{m_X}{2 Q_A} \, (v_{\infty}^2 + \alpha) \right] \, \Theta_{\alpha} 
\, . \label{eq2.19}
\end{eqnarray}
In the second form, simplified by taking $v_{\infty} (r , v_{\rm loc}) = 
(v_{\rm loc}^2 - v_{\rm esc}^2 (r))^{1/2}$ as integration variable, we 
used $v_{\rm loc} \, d v_{\rm loc} = v_{\infty} \, d v_{\infty}$ and 
Liouville's theorem. Note that the step function $\Theta_{\alpha}$ limits 
$v_{\infty}^2$ to the range $0 \leq v_{\infty}^2 \leq \beta_-^A (v_{\rm 
esc}^2 (r) - \alpha / \beta_+^A)$. With the definition (\ref{eq2.19}), 
the result (\ref{eq2.18}) can be finally written as (after integration 
over the Sun)
\begin{equation}
\left. \frac{d \dot{N}_A}{d \, \alpha} \right\vert_{J \geq J_{\min}} 
=  \frac{n_X}{v_o} \int_{r \geq r_{\min}} d^3 {\bf x} \, n_A ( r)
 \, \sigma_A \left( 1 - \frac{J_{\min}^2}{r^2 \, (v_{\rm esc}^2 (r)
- \alpha)} 
\right)^{1/2} K_A (r,\alpha) \, . \label{eq2.20}
\end{equation}
Evidently, the sum over all the (significant) values of $A$ (atomic
number) present in  the Sun must be ultimately performed. Here, the minimum
radius $r_{\min}$  (perhilion distance) is defined in terms of the minimum
angular momentum $J_{\min}$ by 
$r_{\min} \, (v_{\rm esc}^2 (r_{\min})- \alpha)^{1/2} \equiv J_{\min}$. This 
formula yields the result needed for our purpose. Namely, the rate with 
which WIMPs scatter on nuclei with atomic number $A$ to end up into bound 
solar orbits with semi-major axis within $[a , a + da]$ (corresponding to 
$[\alpha , \alpha + d \alpha]$ with $a \equiv G_N \, M_{\odot} / a$), and with 
specific angular momentum $J \geq J_{\min}$. 

The limit $J_{\min} \rightarrow 0^+$ would collect all the WIMPs 
scattered at any point within the Sun, even with very small perihelion 
distances $r_{\min}$, i.e. passing very near the center of the Sun. In 
our subsequent work we shall be interested in the opposite limit 
$J_{\min} \rightarrow (R_S \, v_{\rm esc} (R_S))^-$ corresponding to orbits 
which graze the Sun (radius $R_S$) and barely penetrate it. The usual 
total capture rate is obtained from Eq.~(\ref{eq2.20}) by setting 
$J_{\min} = 0$ and by integrating over $\alpha = GM_{\odot} / a$. We shall be 
interested in values $J_{\min} \simeq R_S \, v_{\rm esc} (R_S)$ and $a 
\sim 1 \, {\rm AU}$, i.e. $\alpha \sim GM_{\odot} / (1 \, {\rm AU}) \sim v_E^2$ 
where $v_E = 29.8 
\, {\rm km/s}$ is the Earth orbital velocity. Note that for such values 
of $a$, $v_{\rm esc}^2 \gg \alpha$ so that we can, with a very good 
approximation, neglect $\alpha$
with respect to  $v_{\rm esc}^2$ both in the square-root factor 
in Eq.~(\ref{eq2.20}) and in the definition of  $r_{\min}$:
$r_{\min} \, v_{\rm esc}(r_{\min}) \simeq J_{\min}$. Further, for
Sun-grazing orbits we can approximate the radial dependence of the
escape velocity by:   $v_{\rm esc}^2(r) \simeq 2  GM_{\odot} / r$. 
Within these approximations, the differential rate 
$[d \dot{N}_A / d\alpha]_{J_{\min}}$ reads
\begin{equation}
\left. \frac{d \dot{N}_A}{d \, \alpha} \right\vert_{J \geq J_{\min}}
\simeq  \frac{n_X}{v_o} \int_{r \geq r_{\min}} d^3 {\bf x} \, n_A ( r)
 \, \sigma_A \left( 1 - \frac{r_{\min}}{r}
\right)^{1/2} K_A (r,\alpha) \, . \label{eq2.22new}
\end{equation}
 Note also that for the values
of $a$ we are interested in, typically
 $\alpha \sim v_E^2 \ll v_{\infty}^2$, so that the
function $K_A (r,\alpha)$ is nearly independent of $\alpha$. 
Let us finally write more explicitly the result (\ref{eq2.19}).
Inserting the explicit (angle-averaged) shifted 
Maxwellian spectrum (\ref{eq2.5}) into Eq.~(\ref{eq2.19}), one can 
express $K_A$ in terms of the error function
\begin{equation}
\chi (x) \equiv \int_0^x dy \, e^{-y^2} \equiv \frac{\pi^{1/2}}{2} \ {\rm 
erf} \, (x) \, . \label{eq2.21}
\end{equation}
To do this one must use as
 integration variable $x \equiv \sqrt{1 + \widehat{a}_A} 
\, v_{\infty} / v_o$ (where $\widehat{a}_A$ is defined below). Setting
\begin{equation}
\widehat{a}_A \equiv \frac{m_X \, v_o^2}{2 \, Q_A} \ , \ \eta_a \equiv 
\frac{1}{\sqrt{1 + \widehat{a}_A}} \, \frac{v_S}{v_o} \, , \label{eq2.22}
\end{equation}
and, with $A(r) > 0$,
\begin{equation}
(A(r))^2 \equiv (1 + \widehat{a}_A) \, \frac{\beta_-^A}{v_o^2} \, \left( v_{\rm 
esc}^2 (r) - \frac{\alpha}{\beta_+^A} \right) \, , \label{eq2.23}
\end{equation}
yields
\begin{equation}
K_A (r,\alpha) = \frac{\exp \left( - \frac{m_X \, \alpha}{2 \, Q_A} 
\right) \, \exp (-\eta_a^2 \, \widehat{a}_A)}{\pi^{1/2} \,
 \beta_+^A \, (1 + \widehat{a}_A) 
\, \eta_a} \, \Bigl[ \chi \, (A(r) - \eta_a) - \chi \, (A(r) + \eta_a) + 2 \, 
\chi \, (\eta_a) \Bigr] \, . \label{eq2.24}
\end{equation}
In the applications below we shall use the standard value for the 
coherence energy $Q_A$ \cite{jungman}
\begin{equation}
Q_A = \frac{3 \, \hbar^2}{2 \, m_A \, R_A^2} \ , \ R_A = 10^{-13} \, {\rm cm} \, 
\left[ 0.3 + 0.91 \left( \frac{m_A}{{\rm GeV}} \right)^{\frac{1}{3}} \right] \, 
. \label{eq2.25}
\end{equation}

When one can neglect both the form factor $(Q_A \rightarrow \infty)$ and 
the relative velocity of the Sun with respect to the galactic halo 
$(\eta_a \rightarrow 0)$, Eq.~(\ref{eq2.24}) simplifies to the form given
in \cite{damkrauss1}
\begin{equation}
K_A (r,\alpha) \vert_{\eta_a = 0}^{Q_A = \infty} = \frac{2}{\pi^{1/2}} \, 
\frac{1}{\beta_+^A} \, \left[ 1 - \exp \left[ - \frac{\beta_-^A}{v_o^2} 
\, \left( v_{\rm esc}^2 (r) - \frac{\alpha}{\beta_+^A} \right) \right] 
\right] \, . \label{eq2.26}
\end{equation}

\section{Getting some of the captured WIMPs out of the Sun}

The scattering events discussed in the previous Section create a 
population of solar-system bound WIMPs, moving (for $a \sim 1 \, {\rm 
AU}$) on 
very elliptic orbits that traverse the Sun again and again . For the 
values of WIMP-nuclei cross sections we shall be mostly interested in 
below (corresponding to effective WIMP-proton cross sections (see section
VII) in the
range 
$4 \times 10^{-42} - 4 \times 10^{-41} \, {\rm cm}^2$), the mean opacity 
of the Sun for orbits with small perihelion distances is in the range 
$10^{-4} - 10^{-3}$. This means that after only $10^3 - 10^4$ orbits 
(i.e. $\sim 10^3 - 10^4 \, {\rm yr}$) these WIMPs will undergo a second 
scattering event in the Sun, making them loose more energy, i.e. binding 
them even more to the Sun. From Eq.~(\ref{eq2.11}) the average energy 
loss per scattering is
\begin{equation}
\langle Q \rangle = \frac{1}{4} \, m_X \, \beta_+^A \, {\bf v}_{\rm 
loc}^2 \simeq \frac{1}{4} \, m_X \, \beta_+^A \, v_{\rm esc}^2 (r) \simeq 
m_X \left( \frac{m_A}{m_X} \right) \, v_{\rm esc}^2 (r) \, , 
\label{eq3.1}
\end{equation}
where we assumed $m_A \ll m_X$.

Compared to the conserved energy before this (second) scattering event, 
$E = - \frac{1}{2} \, m_X \, \alpha$, the mean energy loss (\ref{eq3.1}) 
is quite large, because $\langle v_{\rm esc}^2 (r) \rangle_{\rm Sun} \sim 
(1000 \, {\rm km/s})^2 \gg \frac{1}{2} \, \alpha \sim \frac{1}{2} \, 
v_E^2 \sim \frac{1}{2} \, (30 \, {\rm km/s})^2$. Even assuming a typical 
mass ratio as small as $m_{16} / 1 \, {\rm TeV} \sim 1.6 \times 10^{-2}$ 
does not compensate the large factor $2 \langle v_{\rm esc}^2 \rangle / 
v_E^2 \sim 2 \times 10^3$. The conclusion is that a second scattering 
event would typically bind the WIMP on an orbit of semi-major axis 
significantly smaller than 1 AU. This irreversible process $(\langle Q 
\rangle > 0)$ leads rather quickly (a few $10^3 - 10^4 \, {\rm yr}$) to 
orbits of size comparable to the Sun (at which stage one might need to 
take into account the thermal velocities of the nuclei in the Sun, 
ultimately leading to a near thermal equilibrium between WIMPs and the 
core of the Sun \cite{spergelpress}).

The conclusion is that most of the population of WIMPs considered in the 
previous Section will end up quickly in the core of the Sun (where they 
will ultimately annihilate with each other, thereby generating an 
interesting indirect neutrino signal, see e.g. \cite{jungman}). The only 
way to save some of these WIMPs from this early demise is to consider the 
fraction of WIMPs that have perihelion distances $r_{\rm min}$ in a small 
range near the radius of the Sun $R_S$, say $R_S (1-\epsilon) \leq r_{\rm 
min} \leq R_S$. As we argued in \cite{damkrauss1} focussing
 on such a subpopulation of
WIMPs has two  advantages: (i) they traverse only a small fraction of the mass
of the Sun and  therefore their lifetime on such grazing orbits is greatly
increased, and (ii)  during this time, the gravitational perturbations due to
the planets can build  up and push them on orbits that no longer cross 
the Sun. We now tackle the latter perturbation problem.

We first review in detail some concepts and notation of Hamiltonian dynamics. In
standard  position-momenta variables the Hamiltonian describing the basic 
interaction between a WIMP and the Sun reads $(r \equiv \vert {\bf x} 
\vert)$
\begin{equation}
{\cal H}_S ({\bf x} , {\bf p}) = \frac{1}{2} \, {\bf p}^2 - U(r) \, , 
\label{eq3.2}
\end{equation}
where $U(r) = + \, G_N \int \rho ({\bf x}') \, d^3 {\bf x}' / \vert {\bf x} - 
{\bf x}' \vert$ is the (spherically symmetric) Newtonian potential 
generated by the mass distribution of the Sun. Thanks to the 
equivalence principle, the mass $m_X$ of the WIMP drops out of the 
problem (even when adding the perturbing influence 
of planets). Therefore in Eq.~(\ref{eq3.2}) and below we simplify the 
writing by factoring out $m_X$ of all quantities, i.e. by working 
formally with a unit-mass WIMP. Because of the spherical symmetry of 
Eq.~(\ref{eq3.2}) we can first introduce spherical coordinates $(r,\theta 
, \varphi , p_r , p_{\theta} , p_{\varphi})$, with respect to which the 
problem is separable, and then work with some associated, convenient 
action-angle variables (``Delaunay variables''). The Jacobi 
time-independent action $S_E (r,\theta,\varphi)$ satisfying ${\cal H}_S 
(q_i , \partial S_0 / \partial q_i) = E$, is of the form \cite{landau}
\begin{eqnarray}
S_E (r,\theta,\varphi) &= &S_r (r) + S_{\theta} (\theta) + S_{\varphi} 
(\varphi) \nonumber \\
&& \nonumber \\
&= &\int dr \, \sqrt{2E + 2U (r) - \frac{J^2}{r^2}} + \int d \theta \, 
\sqrt{J^2 - \frac{J_z^2}{\sin^2 \theta}} + J_z \, \varphi \, . 
\label{eq3.3}
\end{eqnarray}
Using $S_E$ we can introduce the usual (action-angle) Delaunay variables, 
traditionally denoted $(L,G,H ; \ell ,g,h)$ \cite{brouwer}. The action 
variables $L$, $G$, $H$ are related to $E$, $J$ and $J_z$ appearing in 
Eq.~(\ref{eq3.3}) through (with $p_r = dS_r / dr$, $p_{\theta} = 
dS_{\theta} / d\theta$, $p_{\varphi} = dS_{\varphi} / d\varphi$)
\begin{mathletters}
\label{eq3.4}
\begin{eqnarray}
H &\equiv &\frac{1}{2\pi} \int_0^{2\pi} p_{\varphi} \, d\varphi = J_z \, 
, \label{eq3.4a} \\
G &\equiv &\frac{2}{2\pi} \int_{\theta_{\rm min}}^{\theta_{\rm max}} 
p_{\theta} \, d\theta = J \, , \label{eq3.4b} \\
L &\equiv &G + \frac{2}{2\pi} \int_{r_{\rm min}}^{r_{\rm max}} p_r \, dr 
= G + \frac{2}{2\pi} \int_{r_{\rm min}}^{r_{\rm max}} dr \, \sqrt{2E + 
2U(r) - \frac{G^2}{r^2}} \, , \label{eq3.4c}
\end{eqnarray}
\end{mathletters}
where the extra factors 2 compensate for the integrations on the 
intervals $[\theta_{\rm min} , \theta_{\rm max}]$, $[r_{\rm min} , r_{\rm 
max}]$, corresponding to half a period for these variables. The angle 
variables (with period $2\pi$) corresponding to $L$, $G$, $H$ are 
respectively denoted $\ell$, $g$, $h$. [Their meaning will be 
discussed further below.] In these variables the Hamiltonian depends only on
$L$  and $G$ (as is clear from Eq.~(\ref{eq3.4c})), ${\cal H}_S = {\cal H}_S 
(L,G)$, so that the general evolution equations,
\begin{mathletters}
\label{eq3.5}
\begin{eqnarray}
\frac{d\ell}{dt} & = & + \frac{\partial {\cal H}}{\partial L} \ , \ 
\frac{dg}{dt} = + \frac{\partial {\cal H}}{\partial G} \ , \ 
\frac{dh}{dt} = + \frac{\partial {\cal H}}{\partial H} \, , 
\label{eq3.5a} \\
&& \nonumber \\
\frac{dL}{dt} & = & - \frac{\partial {\cal H}}{\partial \ell} \ , \ 
\frac{dG}{dt} = - \frac{\partial {\cal H}}{\partial g} \ , \ 
\frac{dH}{dt} = - \frac{\partial {\cal H}}{\partial h} \, , 
\label{eq3.5b}
\end{eqnarray}
\end{mathletters}
tell us, in the case of the problem (\ref{eq3.2}) that the action 
variables $L$, $G$ and $H$ are constant, while, among the angle 
variables, $h$ is constant, but $\ell$ and $g$ evolve linearly in time: 
$\ell = n t + \ell_o$, $g = \dot{\omega} \, t + g_o$. Here, $n \equiv 
2\pi / P$  is the mean angular frequency of the radial motion $(P =$ 
perihelion to perihelion period), and $\dot \omega$ is the mean rate of 
advance of the perihelion.

The central point is the following. A WIMP orbit 
with a generic perihelion distance $r_{\rm min} \lesssim R_S$  will 
undergo a large perihelion precession $\Delta \, \omega \sim 2\pi$ per orbit, 
i.e. $\dot \omega \sim n$, because the potential $U(r)$ {\it within the 
Sun} is very modified compared to the exterior $1/r$ potential leading (by 
``accident'') to the absence of perihelion motion. In other words, the 
trajectory of the WIMP will generically be a fast advancing {\it rosette}. 
This means that {\it both} angles $\ell$ and $g$ are {\it fast variables}. 
When adding in the small perturbing effect of the planets, i.e. when 
considering the total Hamiltonian,
\begin{equation}
{\cal H}_{\rm tot} = {\cal H}_S (L,G) + {\cal H}_p (L,G,H ; \ell ,g,h ; 
L_p , \ldots , \ell_p , \ldots ) \, , \label{eq3.6}
\end{equation}
where ${\cal H}_p$ (which contains a small factor $\mu_p = m_{\rm planet} 
/ M_{\odot}$) is the planetary perturbation, we can work out the (first-order) 
{\it secular} effects due to the planets by averaging over the fast 
variables $\ell$ and $g$ (as well as the mean anomalies $\ell_p$ of the 
planets). Then the first two equations (\ref{eq3.5b}) tell us immediately 
that the corresponding action variables $L$ and $G$ are secularly constant 
because planetary perturbations average out to zero (e.g. $\langle dG / dt 
\rangle = - \langle \partial {\cal H} / \partial g \rangle_{\ell , g} 
\equiv 0$, because of the averaging over the fast angle $g$). As we shall 
see below $L$ is essentially related to the semi-major axis $a$ of the 
WIMP orbit, while $G/L$ is related to the eccentricity $e$. The conclusion 
is that when the rosette motion is fast, planetary perturbations do not 
induce any secular evolution in the semi-major axis and in the 
eccentricity of the WIMP orbit. Therefore such WIMP orbits will 
necessarily traverse the Sun again and again and end up falling in its
core.

A different situation arises for WIMP orbits that graze the Sun.  Throughout
their orbits they feel essentially a $1/r$ potential due to the Sun, so that
their  rosette motion will be very slow. Consequently, the variable $g$ will be 
slow for them (compared to $\ell$), and we cannot average on $g$. To 
tackle this problem we split the total Hamiltonian (\ref{eq3.6}) in 
three parts,
\begin{equation}
{\cal H}_{\rm tot} = {\cal H}_o + {\cal H}_1 + {\cal H}_p \, , 
\label{eq3.7}
\end{equation}
where we take as unperturbed Hamiltonian the one corresponding to a 
point-like Sun, 
\begin{equation}
{\cal H}_o = \frac{1}{2} \, {\bf p}^2 - \frac{G_N \, M_{\odot}}{r} \, , 
\label{eq3.8}
\end{equation}
while the perturbations are
\begin{equation}
{\cal H}_1 = -\delta \, U(r) \ , \ {\cal H}_p = - \sum_p \ G_N \, m_p \left( 
\frac{1}{\vert {\bf x}_X - {\bf x}_p \vert} - \frac{{\bf x}_X \cdot {\bf 
x}_p}{\vert {\bf x}_p \vert^3} \right) \, . \label{eq3.9}
\end{equation}
Here, $\delta \, U(r) \equiv U(r) - \ G_N \, M_{\odot} / r$ is the non $1/r$ 
part of the potential generated by the Sun ($\delta \, U$ is zero when
$r > R_S$ and is responsible for the rosette effect when an orbit
 penetrates the 
Sun), and ${\cal H}_p$ denotes the planetary perturbations \cite{brouwer}. 
It is a sum over the planets with mass $m_p$ and heliocentric positions 
${\bf x}_p$. [The last term comes from the transformation between inertial 
(barycentric) coordinates and heliocentric ones.] We shall henceforth use 
the unperturbed Delaunay variables defined by the Hamiltonian ${\cal H}_o$ 
(corresponding to Keplerian motion). They are explicitly given in terms of 
the usual elliptic elements by
\begin{mathletters}
\label{eq3.10}
\begin{eqnarray}
L & = & \sqrt{G_N \, M_{\odot} \, a} \ , \ G = \sqrt{G_N \, M_{\odot} \, a (1-e^2)} \ 
, \ H = \sqrt{G_N \, M_{\odot} \, a (1-e^2)} \, \cos i \, , \label{eq3.10a} \\
\ell & = & \hbox{mean anomaly}, \ g = \omega = \hbox{periastron argument}, 
\nonumber \\
h & = & \Omega = \hbox{longitude of the ascending mode}. \label{eq3.10b}
\end{eqnarray}
\end{mathletters}
Here, $i$ denotes the inclination (with respect to the ecliptic), and we 
recall that the mean anomaly is, in Keplerian motion the angle $\ell = nt 
+ \ell_o$ where $n = 2\pi / P$ is the radial circular frequency. It will 
be very convenient in this Section to use units such as
\begin{equation}
G_N \, M_{\odot} = 1 \Rightarrow {\cal H}_o = - \frac{1}{2L^2} \ , \ L = 
\sqrt{a} \ , \ G = \sqrt{a (1-e^2)} \ , \ H = G \cos i \, . \label{eq3.11}
\end{equation}
The fact that the unperturbed Hamiltonian depends only on the action 
variable $L = I_r + I_{\theta}$, Eq.~(\ref{eq3.4c}), is the famous 
degeneracy of the Coulomb problem. [In quantum language, $L$ corresponds 
the principal quantum number $n_q = n_r + n_{\theta} = n_r + j$, while 
$G=J$ corresponds to $j$ and, $H = J_z$ to $m$.] It implies immediately 
from the canonical equations (\ref{eq3.5}) that the only fast angle 
variable is $\ell = nt + \ell_o +$ perturbations, where $n \equiv \partial 
{\cal H}_o / \partial L = L^{-3} = a^{-3/2}$. This is just Kepler's law, 
$n^2 \, a^3 = 1$, in our units where $G_N \, M_{\odot} = 1$. 

We are interested 
in deriving the {\it secular} evolution of the elliptic elements $a$, $e$, 
$i$, or equivalently $L$, $G$, $H$, under the combined influence of the 
perturbations ${\cal H}_1$ and ${\cal H}_p$. This is simply obtained by 
averaging the canonical equations over the fast angles, i.e. all the mean 
anomalies of the problem: $\ell$, $\ell_p$. [We denote this average by an 
overbar.] By averaging Eqs.~(\ref{eq3.5b}) one sees easily that (in first 
order) $L$ will be secularly constant (i.e. $a =$ const), while $G$, $H$, 
$g$, $h$ slowly evolve under the averaged perturbed Hamiltonian
\begin{equation}
{\cal H}_{\rm pert} (L,G,H;g) = \overline{\cal H}_1 (L,G) + \overline{\cal 
H}_p (L,G,H;g;L_p) \, . \label{eq3.12}
\end{equation}
As already indicated in Eq.~(\ref{eq3.12}), the averaged perturbed 
Hamiltonian depends (besides $L = \sqrt{a}$ and $L_p = \sqrt{a_p}$ which 
can be treated as constants) only on $G$, $H$ and $g = \omega$. We 
consider, for simplicity, that the planets move on circular orbits. The 
lack of dependence on the last angle $h = \Omega$ comes from the averaging 
over $\ell$ and $\ell_p$ which establishes an azimuthal symmetry in the 
averaged Hamiltonian. Because of this azimuthal symmetry, $dH / dt = 
-\partial {\cal H}_{\rm pert} / \partial h = 0$, i.e. $H = J_z$ is a 
(secular) constant of the motion. Finally, we can treat $L$, $H$ and $L_p$ 
as constants, and consider only the evolution of the canonical pair $(G,g) 
= (J,\omega)$ under the Hamiltonian ${\cal H}_{\rm pert} (G,g) = 
\overline{\cal H}_1 (G) + \overline{\cal H}_p (G,g)$. But this is, in 
principle, a rather simple problem because we have now only one degree of 
freedom (one canonical pair), with a time-independent Hamiltonian. We 
conclude immediately that the perturbed energy is a constant of the 
motion. Therefore, it will be essentially enough to draw the phase-space 
picture of the problem via Hamiltonian level curves,
\begin{equation}
{\cal H}_{\rm pert} (G,g) = \overline{\cal H}_1 (G) + \overline{\cal H}_p 
(G,g) = E_{\rm pert} = {\rm const} \, , \label{eq3.13}
\end{equation}
to be able to control the secular evolution of $G$, i.e. of the angular 
momentum $J$. We must now compute explicitly the values of $\overline{\cal 
H}_1$ and $\overline{\cal H}_p$ to see when planetary perturbations can, 
via the equation $dG / dt = - \partial {\cal H}_{\rm pert} / \partial g = 
- \partial \overline{\cal H}_p / \partial g$, force $G=J$ to evolve 
ultimately towards large enough values of $J$, corresponding to orbits 
which no longer traverse the Sun.

Let us first deal with the planetary perturbations. The calculation of 
$\overline{\cal H}_p$ over the WIMP phase space is, in principle, a 
complicated matter because the WIMP 
can sometimes have a near collision with a planet, especially with Venus 
which is as massive as the Earth and nearer to the Sun. [Remember that we are 
interested in a population of WIMPs with $a \sim 1 \, {\rm AU}$, so that 
they can be 
detected on Earth.] In fact, we tried to estimate separately the effect of 
such near collisions on the evolution of $G=J$. They will cause $G$ to 
undergo a kind of random walk, which can certainly help to increase $G$ 
beyond $J_S = R_S \, v_{\rm esc} (R_S)$. This effect is difficult to 
estimate with the kind of accuracy we attempt below for averaged
perturbations, but it may be comparable to the effect we shall discuss
below. By not including it explicitly, our estimate  below of the
fraction of WIMPs that can get kicked out of the Sun should thus be considered
as a {\it lower bound}. To tackle analytically
$\overline{\cal  H}_p$ we shall assume that the WIMP orbit is sufficiently
smaller than the  planetary orbits considered to be able to expand ${\cal H}_p$
in powers of 
$a/a_p$ and keep only the lowest significant (quadrupolar) contribution. 
[Actually, we have checked numerically that this quadrupolar approximation 
is surprisingly good even up to $a \simeq a_p / 2$ for the very elliptic 
orbits we consider.] Doing the double average over $\ell$ and $\ell_p$ we 
find that the quadrupolar approximation to the second Eq.~(\ref{eq3.9}) 
(actually the last, dipolar, term averages to zero) yields
\begin{eqnarray}
\overline{\cal H}_p &\simeq & -\sum_p \ \frac{1}{2} \, \mu_p \, \langle 
x_X^i \, x_X^j \rangle_{\ell} \, \left\langle \partial_{ij} \, 
\frac{1}{\vert {\bf x}_p \vert} \right\rangle_{\ell_p} \nonumber \\
& = & -\sum_p \ \frac{\mu_p}{4 \, a_p^3} \ \langle r^2 - 3 \, z^2 
\rangle_{\ell} \, . \label{eq3.14}
\end{eqnarray}
In terms of the {\it eccentric anomaly} $u$, we have \cite{brouwer}
\begin{equation}
\ell = u - e \sin u \ , \ r = a \, (1 - e \cos u) \ , \ z = \sin i \, 
(\sin \omega \, \overline x + \cos \omega \, \overline y ) \, , 
\label{eq3.15}
\end{equation}
with ($v$ denoting the true anomaly)
\begin{equation}
\overline x = r \cos v = a \, (\cos u - e) \ , \ \overline y = r \sin v = 
a \, \sqrt{1 - e^2} \sin u \, . \label{eq3.16}
\end{equation}
The average over $\ell$ is easily computed as a modulated average over 
$u$, namely for any quantity $q(\ell) = q(u)$
\begin{equation}
\langle q \rangle_{\ell} = \frac{1}{2\pi} \, \oint d\ell \, q = 
\frac{1}{2\pi} \, \oint du (1 - e \cos u) \, q = \langle (1 - e \cos u) \, 
q \rangle_u \, . \label{eq3.17}
\end{equation}
This yields
\begin{equation}
\left\langle \frac{r^2}{a^2} \right\rangle_{\ell} = 1 + \frac{3}{2} \, e^2 
\ , \ \left\langle \frac{z^2}{a^2} \right\rangle_{\ell} = \frac{1}{2} \, 
\sin^2 i \, [1 + e^2 \, (4-5 \cos^2 \omega)] \, , \label{eq3.18}
\end{equation}
so that
\begin{equation}
\overline{\cal H}_p = \sum_p - \frac{1}{4} \, \mu_p \, \frac{a^2}{a_p^3} 
\, \left[ 1 + \frac{3}{2} \, e^2 - \frac{3}{2} \, \sin^2 i - \frac{3}{2} 
\, e^2 \sin^2 i \, (5 \sin^2 \omega - 1) \right] \, . \label{eq3.19}
\end{equation}
To use this Hamiltonian in the canonical equations (\ref{eq3.5}) one 
should think of all the elliptic elements $a$, $e$, $i$, $\omega$ as being 
expressed in terms of $L$, $G$, $H$, $g$. In particular, we recall 
that $L = \sqrt{a}$ and $H = G \cos i$ can be considered as constants, 
while $G = \sqrt{a (1-e^2)}$ and $g \equiv \omega$ are the evolving 
variables. 
The dynamics resulting from the Hamiltonian 
$\overline{\cal H}_p(G,g)$,
considered separately from $\overline{\cal H}_1(G,g)$, 
i.e. for orbits that stay outside 
the Sun,
was studied some time ago by Kozai \cite{kozai} (see also 
the works of the Russian 
school
in Ref.~\cite{beletsky}, and Refs.~\cite{TM96,HTT97} 
for studies in which this Hamiltonian 
is
relevant).
When considering the beginning of the evolution of $G$ (while 
the WIMP is still in the outskirts of the Sun), it is essentially enough 
to use as Hamiltonian the $g$-dependent part of Eq.~(\ref{eq3.19}), i.e.
\begin{equation}
\overline{\cal H}_p^g = \sum_p + \frac{15}{8} \, \mu_p \, 
\frac{a^2}{a_p^3} \, e^2 \sin^2 i \, \sin^2 g \, , \label{eq3.20}
\end{equation}
with $e \approx 1$ (very elliptic orbits) and $i \approx$ const (because 
the fractional variation of the inclination while the WIMP is still in the 
Sun is small, being correlated through $H = G \cos i =$ const to the 
fractional variation of $G = J$. Indeed, the WIMP just migrates through an 
outer skin of the Sun).

Let us now consider the other perturbation, $\overline{\cal H}_1$, linked 
to the mass distribution in the Sun, Eq.~(\ref{eq3.9}). For this term, it 
is more convenient to write the $\ell$-average in terms of an integral 
over the radial variable $r$ (in terms of which $\delta \, U (r)$ is most 
easily expressed). To do that we can first replace the $\ell$ (or time) 
average by an average over the area measure $r^2 \, dv$ (Kepler's area 
law). Here, $v$ is the true anomaly, so that the unperturbed trajectory on 
which we integrate is the ellipse $r = a \, (1 - e^2) / (1 + e \cos v)$. 
We can here replace $a \, (1 - e^2)$ by $J^2 = G^2$ (in our units) and 
replace $e$ in the denominator (only) by one. [Indeed, we deal with very 
elliptical orbits with $1 - e^2 \sim R_S / a \ll 1$.] Differentiating the 
polar equation of the ellipse (approximated as a parabola) yields
\begin{equation}
r^2 \, dv \simeq \pm \frac{J \, r \, dr}{\sqrt{2r - J^2}} \, . 
\label{eq3.21}
\end{equation}
We then get
\begin{equation}
\overline{\cal H}_1 = - \frac{1}{\pi \, a^{3/2}} \int_{r_{\rm 
min}}^{r_{\rm max}} \frac{r \, dr}{\sqrt{2r - J^2}} \, \delta \, U (r) = - 
\frac{1}{2^{1/2} \, \pi \, a^{3/2}} \int_{r_{\rm min}}^{R_S} \frac{r \, 
dr}{\sqrt{r - r_{\rm min}}} \, \delta \, U (r) \, . \label{eq3.22}
\end{equation}
Here, $r_{\rm min} (J)$ is the perihelion distance (closest radius) of the 
orbit such that $J^2 = 2 \, r_{\rm min} \simeq (r \, v_{\rm esc} (r))^2$, 
consistently with our definition of $r_{\rm min}$ (for $J = J_{\rm min}$) 
in Section 2. [Indeed, $v_{\rm esc}^2 (r) \simeq 2 \, G_N \, M_{\odot} / r$.] 
The maximum radius (which would be formally infinite for the parabolic 
approximation) plays no role because $\delta \, U (r)$ vanishes by 
definition outside the Sun.

To express explicitly $\overline{\cal H}_1$ in terms of $r_{\rm min}$ 
(i.e. of $G \equiv J = \sqrt{2 \, r_{\rm min}}$) we need to know the 
radial dependence of $\delta \, U (r)$ in the actual Sun. To do that we 
have fitted the radial mass distribution of the Sun, given numerically by 
Bahcall and Pinsonneault \cite{bahcall}, to simple power laws. We find 
that the mass distribution in the outskirts of the Sun (we are interested 
in the last outer few percent of the mass of the Sun) is well approximated 
by the fit
\begin{equation}
\frac{\delta \, M(r)}{M_{\odot}} \equiv 1 - \frac{M(r)}{M_{\odot}} \simeq \epsilon_m 
\left[ \left( \frac{\overline{R}_S}{r} \right)^3 - 1 \right] \ ; \ \ 
( r \gtrsim  0.55 R_S) \ , 
\label{eq3.23}
\end{equation}
where
\begin{equation}
\epsilon_m = 0.0219 \ , \ \overline{R}_S = 0.907 \, R_S \, . 
\label{eq3.24}
\end{equation}
When using this fit, one must use $\overline{R}_S$ as an effective radius 
of the Sun beyond which the density vanishes (and $\delta \, U (r) = 0$). 
[This neglects only the outer $0.13~\%$ mass of the actual Sun.] From the 
(effective) radial mass distribution (\ref{eq3.23}) we can deduce both the 
density distribution $(\rho (r) \propto r^{-6} \, \Theta (\overline{R}_S - 
r))$ and (by integration of $d \, U(r) / dr = -G_N \, M(r) / r^2$) the 
Newtonian potential $U(r) = G_N \, M_{\odot} / r + \delta \, U(r)$. This gives
\begin{equation}
\delta \, U(r) = \epsilon_m \, \frac{G_N \, M_{\odot}}{\overline{R}_S} \left[ 
\frac{\overline{R}_S}{r} - \frac{1}{4} \left( \frac{\overline{R}_S}{r} 
\right)^4 - \frac{3}{4} \right] \Theta \, (\overline{R}_S - r) \, . 
\label{eq3.25}
\end{equation}
By inserting (\ref{eq3.25}) into (\ref{eq3.22}) we finally get
\begin{equation}
\overline{\cal H}_1 (J) = \frac{\epsilon_m \, \overline{R}_S^{1/2}}{\pi \, 
2^{1/2} \, a^{3/2}} \, h (x_{\rm min}) \ , \ x_{\rm min} \equiv 
\frac{r_{\rm min}}{\overline{R}_S} \equiv \frac{J^2}{2 \, \overline{R}_S} 
\, , \label{eq3.26}
\end{equation}
where we used $G_N \, M_{\odot} = 1$, and where we defined the dimensionless 
function
\begin{equation}
h (x_{\rm min}) \equiv \Theta \, (1 - x_{\rm min}) \int_{x_{\rm min}}^1 
\frac{dx \, x}{\sqrt{x - x_{\rm min}}} \left[ \frac{3}{4} + \frac{1}{4 \, 
x^4} - \frac{1}{x} \right] \, . \label{eq3.27}
\end{equation}
It is important to realize that as $x_{\rm min} \rightarrow 1$, i.e. as we 
get out of the Sun, the (Hamiltonian) function $h(x_{\rm min})$ tends to 
zero as fast as
\begin{equation}
h (x_{\rm min}) \simeq \frac{8}{5} \, (1 - x_{\rm min})^{5/2} \, \Theta \, 
(1 - x_{\rm min}) \label{eq3.28}
\end{equation}
(because $\delta \, U(r)$ vanishes as $(r - \overline{R}_S)^2$). By 
contrast, we shall see below that the efficiency of the outer skin in 
orbit-capturing WIMPs vanishes only as $(1 - x_{\rm min})^{3/2}$. This 
difference in asymptotic decrease will help in getting more WIMPs 
out of the Sun.

As was explained above, the problem of selecting the values of the WIMP 
angular momentum $J \equiv G$ for which planetary perturbations are strong 
enough to kick the WIMPs out of the Sun can be reduced to studying the level 
curves of the Hamiltonian (using only the crucial $g$-dependent part of 
$\overline{\cal H}_p$, Eq.~(\ref{eq3.20}))
\begin{equation}
{\cal H}'_{\rm pert} = \overline{\cal H}_1 (G) + \overline{\cal H}_p^g 
(G,g) = \beta \, h \left( \frac{G^2}{2 \, \overline{R}_S} \right) + \alpha 
\sin^2 g \, , \label{eq3.29}
\end{equation}
where
\begin{equation}
\alpha = \frac{15}{8} \, a^2 \sin^2 i \, \sum_p \, \frac{\mu_p}{a_p^3} \ , 
\ \beta = \frac{\epsilon_m}{\pi \, 2^{1/2}} \, 
\frac{\overline{R}_S^{1/2}}{a^{3/2}} \, . \label{eq3.30}
\end{equation}
Remember that the Hamiltonian function $h(x_{\rm min})$ (\ref{eq3.27}) 
vanishes for $x_{\rm min} \geq 1$, i.e. when the angular momentum exceeds 
the value corresponding exactly to a grazing incidence, and behaves as 
$(8/5) \, (1 - x_{\rm min})^{5/2}$ when $x_{\rm min} \rightarrow 1^-$. The 
level curves of  (\ref{eq3.29}) for $x_{\rm min} < 1$ have the same 
shape as that of the Hamiltonian $\beta' (G - G_S)^2 + \alpha \sin^2 g$. 
This is nothing but a pendulum: $H_{\rm pend} = \frac{1}{2} \, 
p_{\theta}^2 + \frac{1}{2} \, k \sin^2 \theta$. As is well known an 
essential element of the phase portrait of a pendulum is the {\it 
separatrix}, i.e. the level curve $H_{\rm pend} = \frac{1}{2} \, k$ which 
passes through the unstable equilibrium position $p_{\theta} = 0$, $\theta 
= \pi / 2 \, {\rm mod} \, \pi$. This curve separates the oscillation 
motions (with $\theta$ oscillating, nonharmonically, around $\theta = 0 \, 
{\rm mod} \, \pi$) from the libration ones (with $\theta$ changing 
monotonically, i.e. the pendulum going around in a sling motion). 
Similarly, in the case of the Hamiltonian (\ref{eq3.29}) it is easy to see 
that the level curve ${\cal H}'_{\rm pert} = \alpha$ (passing through $G^2 
= 2 \, \overline{R}_S$, $g = \pi / 2 \, {\rm mod} \, \pi$), i.e.
\begin{equation}
h \left( \frac{G^2}{2 \, \overline{R}_S} \right) = \lambda_{a,i} \cos^2 g 
\, , \label{eq3.31}
\end{equation}
is a separatrix (for $G < G_S \equiv \sqrt{2 \, \overline{R}_S}$). Here
\begin{mathletters}
\label{eq3.32}
\begin{eqnarray}
\lambda_{a,i} & = & \frac{\alpha}{\beta} = \lambda_1 \left( \frac{a}{a_1} 
\right)^{7/2} \sin^2 i \ , \ a_1 \equiv 1 \, {\rm AU} \, , \label{eq3.32a} 
\\
\lambda_1 & = & \frac{15}{8} \, \pi \, 2^{1/2} \, \frac{1}{\epsilon_m} 
\left( \frac{a_1}{\overline{R}_S} \right)^{1/2} \, \sum_p \, \mu_p \left( 
\frac{a_1}{a_p} \right)^3 \, . \label{eq3.32b}
\end{eqnarray}
\end{mathletters}

The separatrix (\ref{eq3.31}) divides between:
(i) the uninteresting (for  us) {\it circulation} motions where the ``momentum''
$p_{\theta} \sim G -  G_S$ keeps the same (negative) sign, i.e. the WIMP
keeps traversing  the Sun over and over again, and (ii) the {\it oscillation}
(or {\it libration})
motions where  the ``momentum'' $p_{\theta} \sim G - G_S$ starts with a
negative value  and evolves so as to reach the value zero (corresponding to a
trajectory  which exactly grazes the Sun) in a finite time. Beyond the boundary
$G =  G_S$ the Hamiltonian describing the evolution of the canonical pair 
$(G,g)$ is the planetary perturbation $\overline{\cal H}_p (L,H;G,g)$ 
given by Eq.~(\ref{eq3.19}) in which one must replace
\begin{equation}
a = L^2 \ , \ e^2 = 1 - \frac{G^2}{L^2} \ , \ \sin^2 i = 1 - 
\frac{H^2}{G^2} \ , \ \omega \equiv g \, , \label{eq3.33}
\end{equation}
where $L$ and $H$ are treated as constants. 
The corresponding phase portrait is drawn in Figure 1. [This Figure
displays the level curves of the more exact perturbation Hamiltonian
(\ref{eq3.13}), with $\overline{\cal H}_p (L,H;G,g)$ given by 
Eq.~(\ref{eq3.19}), instead of the approximate expression (\ref{eq3.20})
used in the text.] The horizontal axis is the angle $g$ (in radians),
while the vertical axis is $\log_{10}(\widehat{G})$, where
$\widehat{G} \equiv G/L = \sqrt{1 - e^2}$ denotes a ``reduced''
angular momentum. The numerical values used in Figure 1 are
$a = 0.844 a_1$, and an initial inclination 
$\cos^2( i_{\rm in}) = 1/3$. The corresponding value for the Sun-grazing
reduced angular momentum is $\widehat{G}_S = 0.1$. The separatrix
discussed above lies in the domain $ \widehat{G} < \widehat{G}_S$, i.e., 
$\log_{10}(\widehat{G}) < -1$, and defines the dividing line (not
explicitly shown) between the trajectories that stay always below
 $\widehat{G}_S$ and those which evolve towards larger  $\widehat{G}$
values.  Note that, starting on the ``initial surface''
$G =  G_S$, most trajectories (except a small neighbourhood of $g = 0 \, {\rm 
mod} \, \pi$) then evolve well away from the Sun to 
undergo large changes of $ \widehat{G}$, up to values of order unity.
 In other words, once the very elliptic WIMP trajectory 
(initially with $ \widehat{G}^2 = 1 - e^2 \sim 
\overline 2 \,{R}_S / a \sim \frac{1}{100}$) exits the Sun, it undergoes, under 
the planetary perturbations, a slow, secular evolution of its eccentricity 
and inclination, up to values $ \widehat{G}^2 = 1 - e^2 \sim 1$
 and corresponding high 
inclinations $i \sim \frac{\pi}{2}$, keeping $H/L = \sqrt{1 - e^2} \cos i$ 
constant.


 We estimated the time scale for exiting the Sun, when starting 
on an initial ``oscillation'' trajectory with $G_{\rm in} < G_S$. From the 
canonical equation $dG / dt = - \partial {\cal H}'_{\rm pert} / \partial 
g$ and the constancy of ${\cal H}'_{\rm pert}$, Eq.~(\ref{eq3.29}), the 
time it takes for $G_{\rm in}$ to evolve up to $G_S$ is given by 
integrating
\begin{equation}
\frac{dt}{dG} = \mp \, (\alpha^2 - 4 \, [\beta \, h (G/G_S) - c ]^2)^{-1/2} 
\, , \label{eq3.34}
\end{equation}
where the constant $c$ is related to the constant energy. By numerically 
integrating (\ref{eq3.34}) over typical trajectories (not too near the 
separatrix on which it takes an infinite time to reach $G=G_S$) we found 
that it generically takes (for $a \sim1$) less than $10^3$ WIMP radial 
periods (i.e. less than $10^3 \, {\rm yr}$) for the eccentricity of the 
WIMP to increase sufficiently to exit the Sun. Then, when $G > G_S$ 
the time scale for the evolution of $G$ is given by the planetary 
perturbations alone and is roughly $[ \sum_{p} \, 
\mu_p (a/a_p)^3 ]^{-1}$ longer than one orbital period, i.e. roughly 
$10^5 \, {\rm yr}$ for $a \sim a_1 \equiv 1 \, {\rm AU}$. After this time, 
the WIMP would, if it evolved only under the simplified planetary 
Hamiltonian $\overline{\cal H}_p$, come back again to low values of $G$, 
corresponding to Sun-penetrating orbits. Under the influence 
of $\overline{\cal H}_1$, it would then again bounce back away from the 
Sun in $\sim 10^3$ orbits. For the scattering cross sections we shall be 
discussing below, the opacity of the small outer skin of the Sun we are 
interested in is typically smaller than $\sim 10^{-6}$. Therefore the 
above process could persist for thousands of cycles before the WIMP gets 
scattered by a nucleus in the outskirts of the Sun. However, as we
mentioned earlier, it is clear 
that the real gravitational interaction of the WIMP with planets is much 
more complicated than what is described by $\overline{\cal H}_p$. In 
particular, the non zero eccentricities of the planets, and the 
occurrence, once in a while, of a near collision with an inner planet will 
cause the elliptic elements of the WIMP to diffuse chaotically away from 
the simplified periodic history described above. Moreover, the very high 
eccentricities (for AU-size orbits) needed to traverse again the Sun 
represent only a very small fraction of the phase space into which the 
WIMP can diffuse.  It thus seems clear that on time scales of several million 
years most of the population of WIMPs we are talking about will have 
irreversibly evolved onto trajectories on which the WIMP can survive 
(without being scattered again in the Sun) for the age of the solar 
system. We will come back later to the problem of the long-term survival 
of such WIMPs on orbits that stay within the inner solar system (rather 
than diffusing out into the outer solar system, and eventually to infinity).

\section{Estimating the capture rate of long-surviving, solar-system bound 
WIMPs}

In view of the previous estimates and arguments, we can consider that the 
population of WIMPs that, after a first scattering event in the Sun, 
diffuse out onto long-surviving solar-bound orbits is given by 
all the initial conditions $G_{\rm in}$, $g_{\rm in}$, $i_{\rm in}$ which 
are ``above" the separatrix (\ref{eq3.31}) (meaning $G_{\rm separatrix} < G 
< G_S$). The logic for quantitatively estimating that population
 is the following. For each given initial 
values of $a$, $g_{\rm in}$ and $i_{\rm in}$, the separatrix defines, by 
solving Eq.~(\ref{eq3.31}) with respect to $G$, a corresponding minimum 
value of $G$
\begin{equation}
J_{\rm min} \equiv G_{\rm min} = G_{\rm separatrix} (a, g_{\rm in} , 
i_{\rm in}) \, , \label{eq4.1}
\end{equation}
such that the trajectories with $J > J_{\rm min}$ end up out of the Sun. 
The differential capture rate corresponding to this class $J > J_{\rm 
min}$ is then precisely defined by our previous result (\ref{eq2.20}) 
which defines
\begin{equation}
\frac{d \, \dot{N}_A}{d \, \alpha} = {\cal C}_A \, [ J_{\rm min} (a, 
g_{\rm in} , i_{\rm in}) , \alpha ] \, . \label{eq4.2}
\end{equation}
Then the actual capture rate is obtained by averaging (\ref{eq4.2}) over 
the distribution of initial values $g_{\rm in}$, $i_{\rm in}$.

Let us first approximate the result (\ref{eq2.20}) by a simpler 
expression. In the radial integration of Eq.~(\ref{eq2.20}) the crucial 
features are the radial dependence of the abundance of element $A$, $n_A 
(r)$, and the square root factor which vanishes at $r = r_{\rm min}$. By 
contrast, the function $K_A (r,\alpha)$, Eq.~(\ref{eq2.24}), varies 
fractionally very little over the small integration range $\overline{R}_S 
(1 - \epsilon) < r < \overline{R}_S$ we are interested in. Therefore a 
good approximation of Eq.~(\ref{eq2.20}) consists of taking out in front a 
factor $K_A (\overline{R}_S , \alpha)$ and performing the radial integration 
on the remaining $r$-dependent factors. To do that let us consider again 
the $r$-dependence of the total mass of the Sun. Denoting $\mu (r) \equiv 
M(r) / M_{\odot}$, with $0 \leq \mu (r) \leq 1$, we have from Eq.~(\ref{eq2.23})
\begin{equation}
d \mu (r) = 3 \, \epsilon_m \, \frac{\overline{R}_S^3 \, dr}{r^4} = 3 \, 
\epsilon_m \, \frac{dx}{x^4} \, , \label{eq4.3}
\end{equation}
where $\epsilon_m$ was given in Eq.~(\ref{eq3.24}) and where $x \equiv r / 
\overline{R}_S$. The distribution of the density of element 
$A$ can be written as
\begin{equation}
d^3 \, {\bf x} \, n_A ({\bf x}) = 4 \pi \, r^2 \, dr \, n_A (r) = f_A \, 
\frac{M_{\odot}}{m_A} \, d \mu (r) \, , \label{eq4.4}
\end{equation}
where $f_A$ denotes the mass fraction of element $A$ in the outskirts of 
the Sun. Let us define the following (``capture'') function
\begin{equation}
c \, (x_{\rm min}) \equiv \int_{x_{\rm min}}^1 \frac{dx}{x^4} \, \sqrt{1 - 
\frac{x_{\rm min}}{x}} = \int_{x_{\rm min}}^1 \frac{dx (x - x_{\rm 
min})^{1/2}}{x^{9/2}} \, . \label{eq4.5}
\end{equation}
In terms of this capture function, the capture rate reads
\begin{equation}
\left. \frac{d \, \dot{N}_A}{d \, \alpha} \right\vert_{J_{\rm min}} = 
\frac{n_X}{v_o} \, f_A \frac{M_{\odot}}{m_A} \, \sigma_A \, K_A^s (\alpha) \, 3 
\, \epsilon_m \, c \, (x_{\rm min}) \, , \label{eq4.n1}
\end{equation}
where $K_A^s$ is the ``surface'' value of the radial function explicitly 
defined in Eq.~(\ref{eq2.24})
\begin{equation}
K_A^s (\alpha) \equiv K_A (\overline{R}_S , \alpha) \, . \label{eq4.n2}
\end{equation}
Note that the surface value of the quantity $A(r)$ entering the error 
functions  in (\ref{eq2.24}) is
\begin{equation}
(A (\overline{R}_S ))^2 \equiv (1 + \widehat{a}_A) \,
 \frac{\beta_-^A}{v_o^2} \left( 
\overline{v}_S^2 - \frac{\alpha}{\beta_+^A} \right) \ , \ \overline{v}_S^2 
= G_N \, M_{\odot} / \overline{R}_S = (648.3 \, \hbox{km/s})^2 \, .
\end{equation}
Actually, as said above, the $\alpha$-dependence of $K_A^s$ is negligible 
(especially in view of all our other approximations) because $\alpha \sim 
v_E^2 \sim (30 \, \hbox{km/s})^2$ so that $\overline{v}_S^2 / \alpha 
\sim 467 \gg 1$. Therefore the $\alpha$-dependence of $d \, \dot{N}_A / 
d\alpha \vert_{J_{\rm min}}$ will come from the $a$-dependence of the 
``ejectable'' radius $x_{\rm min}$, to the estimate of which we now turn. 

Both the ``capture function'' $c \, (x_{\rm min})$ and the previously 
introduced ``hamiltonian function'' $h \, (x_{\rm min})$, 
Eq.~(\ref{eq3.27}), can be explicitly expressed in terms of elementary 
functions. But these explicit expressions will not be really needed here. 
On the other hand, it is important to note the asymptotic behaviour of $c 
\, (x_{\rm min})$ as $x_{\rm min} \rightarrow 1^-$:
\begin{equation}
c \, (x_{\rm min}) \simeq \frac{2}{3} \, (1 - x_{\rm min})^{3/2} \, . 
\label{eq4.6}
\end{equation}
The fact that $c \, (x_{\rm min}) \propto (1 - x_{\rm min})^{3/2}$ 
decreases less fast than $h \, (x_{\rm min}) \propto (1 - x_{\rm 
min})^{5/2}$, Eq.~(\ref{eq3.28}), as $x_{\rm min} \rightarrow 1^-$ is 
important for us because the width of the separatrix (\ref{eq3.31}), $h \, 
(x_{\rm min}) = {\cal O} (\lambda_{a,i})$, will be converted in a capture 
rate proportional to $\lambda_{a,i}^{3/5}$, i.e. something larger than the 
{\it a priori} expected small perturbation parameter $\lambda_{a,i} = 
\alpha / \beta \propto \mu_p / a_p^3$. If we were to use only the 
asymptotic expressions (\ref{eq3.28}), (\ref{eq4.6}) the width of the 
separatrix (\ref{eq3.31}) would be $1 - x_{\rm min} \simeq (5 \, \lambda_1 
/ 8)^{2/5} \, (a/a_1)^{7/5} \, (\sin i)^{4/5} \, (\cos g)^{4/5}$, and the 
corresponding capture rate would be proportional to
\begin{equation}
c_{\rm asymptotic} (x_{\rm min}) \simeq \frac{2}{3} \left( \frac{5 \, 
\lambda_1}{8} \right)^{3/5} \left( \frac{a}{a_1} \right)^{\frac{21}{10}} 
\, (\sin i)^{6/5} \, (\cos g)^{6/5} \, . \label{eq4.7}
\end{equation}
The actual capture rate will be larger than the one predicted by 
(\ref{eq4.7}) because the actual function $c \, (x_{\rm min})$ increases 
faster than (\ref{eq4.6}) as one gets into the Sun. To combine some 
adequate numerical accuracy with the convenience of having analytical 
expressions we shall assume the approximate validity of the scalings in 
$a/a_1$, $\sin i$ and $\sin g$ predicted by Eq.~(\ref{eq4.7}) but 
calculate the precise numerical coefficient applicable for $a/a_1 = \sin i 
= \cos g = 1$ by using the full numerical expressions of the functions 
$h(x)$ and $c \, (x)$, i.e. by inverting $h (x_1) = \lambda_1$ in $x_1$ 
and computing $c \, (x_1)$. To do this we need the numerical value of 
$\lambda_1$. First, taking into account the most significant planets, i.e. 
Venus, the Earth, Mars, Jupiter and Saturn, we find
\begin{equation}
\sum_p \, \mu_p \left( \frac{a_1}{a_p} \right)^3 = 1.67 \times 10^{-5} \, 
, \label{eq4.8}
\end{equation}
where we recall that $a_1$ denotes simply the basic unit for semi-major 
axes, namely the astronomical unit (AU). The other important numerical 
ingredients in $\lambda_1$, Eq.~(\ref{eq3.32b}), are
\begin{equation}
\left( \frac{a_1}{\overline{R}_S} \right)^{1/2} = (236.9)^{1/2} = 15.39 \ 
, \ \epsilon_m = 0.0219 \, , \label{eq4.9}
\end{equation}
so that
\begin{equation}
\lambda_1 = 0.0978 \, . \label{eq4.10}
\end{equation}

The corresponding solution of $h (x_1) = \lambda_1$ is $x_1 = 0.729$ 
(which means that we are typically dealing with the outer $27~\%$ of the 
Sun radius-wise, containing in fact only $\sim 2~\%$ of the mass of the 
Sun). The corresponding value of the capture function is $c \, (x_1) = 
0.172$, so that the numerical combination effectively appearing in the 
capture rate (\ref{eq4.n1}) will equal (when $a/a_1 = \sin i = \cos g = 
1$)
\begin{equation}
3 \, \epsilon _m \, c \, (x_1) = 0.0113 \, . \label{eq4.11}
\end{equation}

This value must, according to Eq.~(\ref{eq4.7}), be scaled by 
$(a/a_1)^{2.1} \, (\sin i)^{6/5} \, (\cos g)^{6/5}$. Here, $i$ and $g$ are 
the initial values of the inclinations and perihelion argument. These 
quantities are random variables: $g$ is expected to have a uniform 
distribution over $[0,2\pi]$, while it is $\cos i$ which is expected to 
have a uniform distribution over $[0,1]$ (indeed, the direction of the 
vectorial angular momentum ${\bf J}$ is expected to be random on the 
celestial sphere). The averaging over these variables brings a factor
\begin{equation}
\langle (\sin i)^{6/5} \rangle_{\cos i} \, \langle (\cos g)^{6/5} 
\rangle_g = 0.7567 \times 0.6007 = 0.4545 \, . \label{eq4.12}
\end{equation}
Together with (\ref{eq4.11}) and the $a$-scaling we end up with a fraction 
kicked out of the Sun,
\begin{equation}
\phi \, (a) \equiv 3 \, \epsilon_m \, \langle c \, (x_{\rm min})
 \rangle \simeq 
\phi_1 \left( \frac{a}{a_1} \right)^{\frac{21}{10}} \ , \ \phi_1 \simeq 
5.13 \times 10^{-3} \, . \label{eq4.13}
\end{equation}
Finally, if we define the $A$-dependent combination
\begin{equation}
g_A \equiv \frac{f_A}{m_A} \, \sigma_A \, K_A^s \, , \label{eq4.14}
\end{equation} 
the rate (per $\alpha = G_N \, M_{\odot} / a$) of solar capture of WIMPs that 
subsequently survive out of the Sun to stay within the inner solar system, 
reads
\begin{equation}
\left. \frac{d \, \dot{N}_A}{d \, \alpha} \right\vert_{\rm surv} = \phi_1 
\left( \frac{a}{a_1} \right)^{\frac{21}{10}} M_{\odot} \, \frac{n_X}{v_o} \, g_A 
\, . \label{eq4.15}
\end{equation}
Note that the $A$-dependence is entirely contained in the quantity $g_A$ 
with dimensions $[\hbox{cross section}] / [\hbox{mass}]$, e.g. $[{\rm 
cm}^2] / [{\rm GeV} / c^2]$, or ${\rm GeV}^{-3}$  in 
particle units. The total capture rate is given by ${\displaystyle \sum_A} 
\, g_A$.

\section{Estimating the present local phase-space distribution of 
surviving solar-captured WIMPs}

The last result (\ref{eq4.15}) of the previous Section gives the rate with 
which a fraction of the WIMP-Sun-scattering events populates a class of 
WIMPs that get out of the Sun with initial semi-major axis equal to $a$, 
and very high initial eccentricities such that
\begin{equation}
1 - e^2 = 2 \, \frac{\overline{R}_S}{a} \simeq 8.44 \times 10^{-3} \, 
\frac{a_1}{a} \, . \label{eq5.1}
\end{equation} 

The question that remains is the following: what happens to 
this 
population while it slowly builds up during the 4.5 Gyr lifetime of 
the 
solar system? What is the present local distribution (in position and 
velocity space) as seen from the Earth of these WIMPs? These questions are 
very difficult to answer with precision because of the complexity of 
Hamiltonian dynamics in the solar system. One would need long-term 
numerical simulations to give reliable quantitative answers. However,
we shall attempt here to make some estimates that will allow
us to estimate the present observable effects of this population of WIMPs.

There are two main worries about the long-term survival of this 
population. 
The first would be that they traverse the Sun again and again and 
end up getting 
accreted by it. We argued away this worry above ( because of the very 
small opacity of the outer skin of the Sun, of the repelling effect of the 
interaction with $\delta \, U(r)$, and of the small probability, given 
some 
additional chaos, that the WIMP again encounters the Sun). Note that 
existing asteroid simulations (see e.g. \cite{gladman}) do not help in this
respect because: (i)  they 
restrict themselves to essentially planar initial data (while our WIMPs 
have fully three-dimensional trajectories), and (ii) they stop their 
numerical integrations as soon as an asteroid touches the Sun (while our 
WIMPs could survive $\sim 10^5 - 10^6$ passes in the outskirts of the 
Sun). [Note that, in the case of the ``lowest" ($a = 2.1 $AU) 
orbits considered in 
\cite{gladman}, $79 \%$ of them were eliminated because of impacting on the 
Sun.]
A second worry concerns the possibility that the adiabatic invariant $a$ 
slowly evolves, in a quasi-random-walk, under the effect of near collisions 
with planets. This diffusion in $a$-space could lead a fraction of the 
WIMPs to have higher and higher values of $a$, possibly being ejected 
from the solar system. One can give a crude analytical estimate of the 
time 
scale on which $a$ can change in the following way. Because of the 
exponential accuracy with which adiabatic invariants are conserved in 
absence of near collisions (i.e. in absence of near singularities in the 
complex plane, see e.g. \cite{landau}), the cause of the random walk of 
$a$ 
must be the existence of near collisions with some planet. Therefore, this 
effect will depend very much on the value of $a$. If $a$ is smaller than 
$a_J / 2 \simeq 2.6 \, a_1$ (where the label $J$ stands for Jupiter) the 
WIMP orbit cannot cross Jupiter's orbit even if the eccentricity
 is very high. In 
that case, only one of the inner planets can have a near collision with a 
WIMP. Let generally $m_p$ be the mass of a planet whose orbit can cross 
that of a WIMP, with semi-major axis $a_X$. The only small parameter in 
the 
problem of the non adiabatic evolution of $a_X$ is $\mu_p \equiv m_p / 
M_{\odot}$ (we use units such that $G_N M_{\odot} = 1$).
 This non adiabatic evolution will be due to a more or less random 
succession of near collisions with the planet. Each collision would induce 
a velocity change $\delta \, v \sim \mu_p / (b \, v)$ and an energy change 
$\delta \, a_X / a_X \sim \pm \mu_p \, a_X / b$ where $b$ is the impact 
parameter. The rate of occurrence of such collisions is smaller as $b$ 
decreases. Between two such quasi-collisions all the angular variables in 
the problem (which determine the $\pm$ sign in the energy change) have 
probably had the chance of being essentially randomized. After a long time 
$t$ we can then consider that the total fractional change of $a_X$, 
$\Delta 
\, a_X$, is a random walk so that one must consider $(\Delta \, a_X / 
a_X)^2 \sim {\displaystyle \sum_{\rm collisions}} (\mu_p \, a_X / b)^2$. 
If 
we provisionally use units where $a_X = 1$, one can see that the typical 
time between two near collisions with impact parameter $b$ is $t_b \sim 
b^{-2}$ ($b^2$ is an effective cross section around the planet, while the 
WIMP is evolving in a full $3D$-volume $\sim a_X^3 = 1$). The number of 
terms in the above random-walk is then $N \sim t / t_b \sim b^2 \, t$ so 
that $(\delta \, a_X)^2 \sim N \, (\mu_p / b)^2 \sim \mu_p^2 \, t$. 
Returning to ordinary units we find a typical diffusion law
\begin{equation}
\left( \frac{\Delta \, a_X}{a_X} \right)^2 \sim \frac{t}{t_D} \ , \ t_D = 
C 
\, \mu_p^{-2} \, T_X \, , \label{eq5.2}
\end{equation}
where $T_X$ is the orbital period corresponding to $a_X$, and where $C$ is 
some numerical constant of order unity. The dimensionless constant $C$ is 
impossible to estimate with any accuracy on the basis of the previous
rough argument (it can also contain  a logarithm
 due to the integration over a relevant range of values of
$b$).   The estimate (\ref{eq5.2}) suggests that the semi-major axis of 
Earth-crossing WIMPs diffuse on a very long time scale $t_D \sim C \, (3.3 
\times 10^5)^2 \, (a_X / a_1)^{3/2} \, {\rm yr} \sim C \times 10^{11} \, 
(a_X / a_1)^{3/2} \, {\rm yr}$, so that we can essentially neglect the 
variation of $a_X$ over the age of the Sun. By contrast, the situation 
becomes dramatically different when $a_X = a_J / 2 = 2.6 \, a_1$, because 
in this case it can cross the orbit of Jupiter with $\mu_p \simeq 
(1047)^{-1}$. This leads to a much shorter diffusion time scale $t_D^J 
\sim 
4.6 \times C \times 10^6 \, (2 \, a_X / a_J)^{3/2} \, {\rm yr}$. The 
existence of such very different time scales depending on $a_X < a_J / 2$ 
or $a_X > a_J / 2$ is well known in asteroid research and is apparent in 
the results of long-term numerical simulations, see, e.g., 
Ref.~\cite{morbidelli}. In principle such numerical simulations can give 
estimates of the diffusion times. It seems that a value of $C \sim 0.1$ is 
roughly compatible with several results and numbers in the literature 
\cite{morbidelli}, \cite{greenberg}, though the comparison might be 
difficult because asteroid simulations start with quasi-planar initial 
conditions.

To summarize: we expect, in first approximation, that the initial 
$a$-distribution derived in previous Sections can build up over $t_S = 4.5 
\, {\rm Gyr}$ with only small diffusion effects if $a < a_J / 2 = 2.6 \, 
a_1$, while if $a > a_J / 2$ this population is cut-off because of a fast 
diffusion in the outskirts of the solar system.
 As we said above, this conclusion, based on 
our analytical
estimate Eq.~(\ref{eq5.2}), should be checked by dedicated long-term numerical
simulations.  

Having discussed the 
secular evolution of $a$, we need now to discuss that of $e$ and the other 
elliptic elements.
As shown in Eq.~(\ref{eq5.1}) the initial values of the eccentricities are 
very near 1. The discussion of the phase portrait of the secular planetary 
Hamiltonian $\overline{\cal H}_p$ in Section III showed that $e$ and $i$ 
undergo large oscillations with $e$ evolving between values very near 1 
and 
values of order one (say 0.3). The lower values of $e$ depend on the value 
of $H = J_z = \sqrt{a (1-e^2)} \cos i$ which is a secular invariant.
Note that we now consider the dynamics implied 
by the Hamiltonian
$\overline{\cal H}_p$, without the solar contribution $\overline{\cal H}_1$. The
phase  portrait of $\overline{\cal H}_p$ differs
from Fig.  1 in that the level curves escaping from the Sun and formerly librating 
around 
$g= 0$ or $\pi$, are now circulating, i.e. such that the angle $g$ is
monotonically  evolving . The phase
portrait of this pure Kozai Hamiltonian is represented, for instance, in
\cite{kozai}  ( Fig. 8),
or \cite{HTT97} (Fig. 3). For our case (a very small value of the constant 
of motion 
$H/L$) this means 
that most orbits will feature a small value of $G/L$, i.e. 
a large value of the 
eccentricity $e$, for
a wide range of values of the periastron argument $g$  
( a maximal eccentricity
is  reached for
$g = \pi/2$ or $3\pi/2$, but this maximum is broad, and large eccentricities
are  maintained during
most of the evolution).

In addition, the time-averaged probability for the eccentricity to fall in the
range
$e 
\pm \frac{1}{2} \, de$, will be peaked toward the extremal values of 
$e(t)$ 
(because $dt = de / \dot e$ diverges there). The maximal value of $e$ is 
very near 1 for all the WIMPs of the population, while the minimal value 
varies across the WIMP population, because it depends, among other initial 
data, on the value of the constant of motion $H = \sqrt{a (1 - e^2)} \cos 
i$. Therefore the overall time-averaged and population-averaged 
distribution function for $e$ will have a peak only near $e \approx 1$. 
But, as we said above,
this peak should be superposed onto a rather flat distribution favoring high 
values of the eccentricity.
( Therefore the fact that the maximal eccentricities of each orbit are reached 
for $g= \pi/2$ or $3\pi/2$,
which implies a lack of spherical symmetry of the spatial distribution of the 
WIMPS, and which thereby somewhat
disfavors the ecliptic plane, should not be numerically very significant).
In 
the absence of detailed numerical simulations of the long-term evolution of 
our 
population of WIMPs this argument (based on the simplified secular 
Hamiltonian (\ref{eq3.19})) suggests to use as an educated guess, for the mean
distribution function of $e$, a distribution  peaked at $e=1$,
i.e. simply a delta function $\delta \, (1-e)$. Numerical  simulations of
asteroids, which go beyond the simplified Hamiltonian  (\ref{eq3.19}) and take
into account near collisions with the planets,  suggest that the
analytically-expected large oscillations in $e$ may  actually be damped and
lead to a population always having quite  large 
eccentricities (see Fig.~4 in Ref.~\cite{morbidelli}). Resolving this 
question, and investigating the actual deviation from 
spherical symmetry of the 
WIMP population, 
would demand the running of long-term numerical simulations.
In  the present paper, we shall assume, for illustration purposes, a 
spherically symmetric population 
formally entirely concentrated at $e \approx 1$,
which is technically simpler to deal with than
a  more realistic 
range of high values of $e$. Certainly, some of the 
details of the predictions that we shall sketch below depend on this 
assumption, but we think that it is an appropriate approximation  at this 
stage, though 
one which will have to be
checked by dedicated numerical simulations.

Under this assumption we can compute both the present space distribution 
and the present velocity distribution of our population of WIMPs. Let us 
first consider the spatial (numerical) density of WIMPs $n (r)$. Consider 
first a subpopulation with some given values of $a$ and $e$. By 
differentiating $r = a \, (1 - e \cos u)$, $\ell = u - e \sin u$ we find
\begin{equation}
d\ell / dr = \pm \, a^{-2} \, r \, (e^2 - (1 - r/a)^2 )^{-1/2} \label{eq5.n1}
\end{equation}
so that the fraction of time, or elementary probability $dp = 2 \vert d\ell / dr 
\vert \, dr / 
2\pi$ (where the extra factor 2 comes from the sign ambiguity), spent by 
this subpopulation within the radii $r$ and $r+dr$ is
\begin{equation}
dp = \frac{1}{\pi \, a^2} \, \frac{r \, dr}{\sqrt{e^2 - (1 - r/a)^2}} \, 
\Theta \, (r - a \, (1-e)) \, \Theta \, (a \, (1+e) - r) \, . 
\label{eq5.3}
\end{equation}

{}From our previous arguments the number of WIMPs with semi-major axis 
within 
$[a,a+da]$ is
\begin{equation}
\frac{dN}{da} \, da = \frac{dN}{d\alpha} \, d\alpha = t_S \, \frac{d \dot 
N}{d\alpha} \, d\alpha \, , \label{eq5.4}
\end{equation}
where $t_S \simeq 4.5 \, {\rm Gyr}$ is the age of the Sun, and where, from 
Eq.~(\ref{eq4.15})
\begin{equation}
\frac{d \dot N}{d\alpha} = \Theta \left( \frac{1}{2} \, a_J - a \right) \, 
\phi_1 \left( \frac{a}{a_1} \right)^{2.1} \, M_{\odot} \, \frac{n_X}{v_o} \, 
g_{\rm tot} \ , \ g_{\rm tot} = \sum_A \, g_A \, . \label{eq5.5}
\end{equation}
The average number of WIMPs within the radii $r$, $r+dr$ is obtained by 
multiplying (\ref{eq5.3}) and (\ref{eq5.4}) and integrating over $a$,
\begin{equation}
d_r \, N = 4 \, \pi \, r^2 \, dr \, n(r) = \int da \, \frac{dN}{da} \, dp 
\, , \label{eq5.6}
\end{equation}
so that the density of WIMPs reads (using $\vert d\alpha \vert = G_N \, 
M_{\odot} 
\, da / a^2$)
\begin{equation}
n(r) = \frac{1}{4 \, \pi^2 \, r} \int da \, \frac{G_N \, M_{\odot}}{a^2} \, t_S 
\, \frac{d \dot N}{d\alpha} \, \frac{\Theta \, (r - a \, (1-e)) \ \Theta 
\, 
(a \, (1+e) - r)}{a^2 \, \sqrt{e^2 - (1 - r/a)^2}} \, . \label{eq5.7}
\end{equation}
If we were to consider a population distributed in eccentricity with 
weight 
$\varphi (e) \, de$, we would need to add a further integration $\int de  
\, \varphi (e)$ in front of Eq.~(\ref{eq5.7}). Here, as we have mentioned, we
shall simply  assume 
that $e \approx 1$ for the entire population. It is then convenient to 
replace $a$ by the new integration variable $x \equiv 2 \, a / r$. 
Inserting Eq.~(\ref{eq5.5}) in Eq.~(\ref{eq5.7}) and remembering the 
various step functions that limit the $a$-integration range we get
\begin{equation}
n(r) = \nu_1 \, n_X \left( \frac{a_1}{r} \right)^{1.9} \, I_n \left( 
\frac{a_J}{r} \right) \, , \label{eq5.8}
\end{equation}
where
\begin{equation}
\nu_1 = \frac{\phi_1}{2^{2.1} \, \pi^2} \, t_S \, \frac{G_N \, M_{\odot}}{a_1^4} 
\, \frac{M_{\odot} \, g_{\rm tot}}{v_o} \, , \label{eq5.9}
\end{equation}
\begin{equation}
I_n \, (y) \equiv \int_1^y \frac{dx}{x^{0.9} \, \sqrt{x - 1}} \, . 
\label{eq5.10}
\end{equation}

As $I_n \, (a_J / r)$ tends to a finite limit as $r \ll a_J$, we see that 
the radial dependence of the WIMP density is essentially given by the 
factor $(a_1 / r)^{1.9}$. This indicates that, if it were possible to do 
so, it would be easier to detect the WIMP population we are talking about 
nearer to the Sun, e.g. by building a detector in a mine (or examining
WIMP-induced tracks in ancient mica) on 
Mercury!  
Let us also note that the radial distribution $n(r)$ probably becomes cut 
off below some radius $r_c < a_M = 0.387 \, a_1$, $M$ being a label for 
Mercury. Indeed, for $a_X < a_M / 2$ WIMPs do not have near collisions 
with 
any of the planets. Their secular orbital evolution should be rather well 
described by the quadrupolar Hamiltonian $\overline{\cal H}_p$, which 
means 
that they will episodically but repeatedly penetrate the outskirts of the 
Sun, thereby risking more to be scattered again. There should exist a 
critical semi-major axis $a_c$, between $\overline{R}_S$ and $a_M / 2$, 
such that when $a_X < a_c$ the WIMP penetrates the Sun too often and
finally gets accreted.

In the following we shall focus on the value of the density of WIMPs 
at the orbital radius of the Earth, $r_E = a_1 \equiv 1 \, {\rm AU}$. Let 
us define the enhancement in WIMP density due to the secondary population 
considered here as
\begin{equation}
\delta_E \equiv \frac{n (a_1)}{n_X} \equiv \frac{\hbox{(secondary) WIMP 
density at the Earth}}{\hbox{halo WIMP density at infinity}} \, . 
\label{eq5.11}
\end{equation}
{}From Eqs.~(\ref{eq5.8}), (\ref{eq5.9}) one finds
\begin{equation}
\delta_E = \phi_2 \, \Delta \, g_{\rm tot} \, , \label{eq5.12}
\end{equation}
with (using $G_N \, M_{\odot} / a_1 = v_E^2$ with $v_E = 29.78 \, {\rm km/s}$, 
and $I_n \, (5.2) = 2.3474$)
\begin{equation}
\phi_2 = \phi_1 \, \frac{I_n \, (5.2)}{2^{2.1} \, \pi^2} = \phi_1 \times 
0.0555 = 2.85 \times 10^{-4} \, , \label{eq5.13}
\end{equation}
\begin{equation}
\Delta = \frac{v_E^2}{v_o} \, t_S \, \frac{M_{\odot}}{a_1^3} = \frac{1.91 
\times 
10^{40}}{(v_o / 220 \, {\rm kms}^{-1})} \, {\rm GeV} \, {\rm cm}^{-2} = 
\frac{7.44 \times 10^{12}}{(v_o / 220 \, {\rm kms}^{-1})} \, ({\rm 
GeV})^3 
\, . \label{eq5.14}
\end{equation}
Finally the local enhancement in density is
\begin{equation}
\delta_E = \frac{5.44 \times 10^{36}}{(v_o / 220 \, {\rm kms}^{-1})} 
\times 
g_{\rm tot} \, {\rm GeV} \, {\rm cm}^{-2} = \frac{0.212}{(v_o / 220 \, 
{\rm kms}^{-1})} \, g_{\rm tot}^{(-10)} \, , \label{eq5.15}
\end{equation}
where $g_{\rm tot}^{(-10)} \equiv 10^{10} \, g_{\rm tot} ({\rm GeV})^3$. 
The meaning of the $10^{10} \, ({\rm GeV})^3$ factor is that in $g_{\rm 
tot} = {\displaystyle \sum_A} \, (f_A / m_A) \, \sigma_A \, K_A^s$, with 
dimensions $[\hbox{mass}]^{-1} \times [\hbox{cross section}]$, one must 
express the mass in units of GeV and the cross section in units of 
$10^{-10} \, {\rm GeV}^{-2}$ (with $\hbar = c =1$). Note the conversion 
factor
${\rm GeV}^{-2} = 3.8938 \times 10^{-28} \, {\rm cm}^2 \, $. 
We shall see in the next Section that $g_{\rm tot}^{(-10)}$ can be 
higher than $\sim 1$, so that this new population could represent a 
significant increase above the standard halo WIMP density.

\section{Observable signals from the new WIMP population}

The secondary WIMP population discussed here could give rise to observationally
significant effects that have not traditionally been taken into account in the
standard approach to dark matter detection, where one considers only the
primary  galactic halo population. The main observable signals from the new 
population are: (i) a new component, involving $\sim {\rm keV}$ energy 
transfer, in the differential spectrum of direct detectors of WIMPs, (ii) a
significantly different angular spectrum in any detector with directional
sensitivity, and and  (iii) a possible significant increase in the indirect
neutrino signal caused by WIMP  annihilations in the Earth. To discuss the
figures of merit associated with the new WIMP population, we need to fold in the
velocity distribution of  the WIMPs. Eq.~(\ref{eq5.n1}) above, together with
$d\ell / dt = n = (G_N 
\, M_{\odot}/ a^3)^{1/2}$, shows that the radial velocity $v_r = dr / dt$ of a 
WIMP passing at radius $r$ reads
\begin{equation}
v_r = \pm \frac{1}{r} \left( \frac{G_N \, M_{\odot}}{a} \right)^{1/2} (e^2 \, 
a^2 - (a-r)^2 )^{1/2} \, . \label{eq5.16bis}
\end{equation}

The local velocity of a WIMP in the Earth frame is ${\bf v}_{\rm loc} = 
{\bf v}_X - {\bf v}_E$, where ${\bf v}_X$ is the heliocentric WIMP 
velocity, whose radial component is (\ref{eq5.16bis}). In our 
approximation where $e \approx 1$ for all the WIMPs ${\bf v}_X$ is in the 
radial direction (with $\vert {\bf v}_X \vert = \vert v_r \vert$), and 
therefore orthogonal to the Earth orbital velocity ${\bf v}_E$. This 
yields $v_{\rm loc}^2 = v_r^2 + v_E^2$ so that, with 
$r = a_1 = 1 \, {\rm AU}$,
\begin{equation}
v_{\rm loc} = v_E \left( 3 - \frac{a_1}{a} \right)^{\frac{1}{2}} \, . 
\label{eq5.17}
\end{equation}
With $ a_1 / 2 \leq a \leq a_J / 2 = 2.6 \, a_1$, this predicts
that the local velocity of the secondary WIMPs we  discuss ranges only between
$v_E = 29.8 \, {\rm km/s}$ and $\sqrt{3 -  1/2.6} \, v_E = 48.2 \, {\rm km/s}$.
These numbers depend on our approximation $e \approx 1$.
They are, however, indicative  of the values we might expect for the actual
population. The local distribution is obtained by eliminating $a$
using (\ref{eq5.4}) and  (\ref{eq5.17}). 

Actually, the observable of most
interest is the  differential rate (events per kg per day and per keV) of
scattering events  in a laboratory sample made of element $A$ \cite{jungman}
\begin{equation}
\frac{dR}{dQ} = \frac{\sigma_A \, n}{2 \, m_{\rm red}^2 (X,A)} \, F_A^2 
(Q) \int_{v_{\rm min} (Q)}^{\infty} \frac{d \, \widehat{n} \, (v)}{v} \ , 
\ v_{\rm min} (Q) = \left( \frac{Q \, m_A}{2 \, m_{\rm red}^2 (X,A)} 
\right)^{1/2} \, . \label{eq5.18}
\end{equation}
Here, $Q = {\bf q}^2 /( 2 \, m_A) = (m_{\rm red}^2 / m_A) \, v_{\rm loc}^2 
(1 - \cos \theta_{\rm cm})$ is the energy transfer from the WIMP $X$ to 
the nucleus $A$, $m_{\rm red} (X,A) \equiv m_X \, m_A / (m_X + m_A)$ is 
the reduced mass, $n$ is the local number density of the considered 
WIMP population, $F_A^2$ the form factor (\ref{eq2.7}), $v$ the local WIMP 
velocity and $d \, \widehat{n} \, (v)$ the {\it normalized} speed 
distribution of the WIMP number density, with $\int d \, \widehat n (v) = 1$. 
Note that for the standard WIMP population $n^{\rm standard} = n_X$, while for 
the new population discussed here $n^{\rm new} = \delta_E \, n_X$. Following 
Ref.~\cite{jungman}, one can introduce for any WIMP population the dimensionless 
quantity
\begin{equation}
T(Q) = \frac{\sqrt{\pi}}{2} \, v_o \int_{v_{\rm min} (Q)}^{\infty} \frac{d \, 
\widehat{n} \, (v)}{v} \, . \label{eq5.19}
\end{equation}
For the standard galactic halo WIMPs
\begin{equation}
T_{\rm standard} (Q) \simeq \exp \, (-v_{\rm min}^2 / v_o^2) = \exp \left( - 
\frac{m_A \, Q}{2 \, v_o^2 \, m_{\rm red}^2} \right) \, , \label{eq5.20}
\end{equation}
when neglecting the motion of the Earth with respect to the halo. For the low 
velocity WIMPs we are considering $T_{\rm standard} (Q) \simeq 1$. Therefore the 
ratio
\begin{equation}
\rho (Q) \equiv \frac{(dR / dQ)^{\rm new}}{(dR / dQ)^{\rm standard}} \equiv 
\delta_E \, \frac{T^{\rm new} (Q)}{T^{\rm standard} (Q)}
 \simeq \delta_E \, T^{\rm new} (Q) \, . \label{eq5.21}
\end{equation}
This quantity is the figure of merit of most interest to us. It expresses the 
fractional increase, with respect to standard expectations, in the differential 
scattering rate. The distribution  $\frac{d \, \widehat{n}^{\rm new}}{da} \, da$ 
is obtained by taking the integrand of Eq.~(\ref{eq5.7}) and normalizing it to 
one. Changing the integration variable from $a$ to $x = 2a / r = 2a / a_1$ gives
\begin{equation}
d \, \widehat{n} = \frac{1}{I_n (5.2)} \, \frac{dx \, \Theta (x-1) \, \Theta 
(5.2 - x)}{x^{0.9} \, \sqrt{x-1}} \, . \label{eq5.22}
\end{equation}
One must then bring in the factor $1/v = 1/(v_E \, \sqrt{3 - 2 x^{-1}})$ from 
Eq.~(\ref{eq5.17}). Let us define the energy scale
\begin{equation}
Q_E \equiv 2 \, \frac{m_{\rm red}^2}{m_A} \, v_E^2 = 2 \left( \frac{m_X}{m_X + 
m_A} \right)^2 \, m_A \, v_E^2 \, . \label{eq5.23}
\end{equation}
For natural Germanium $\langle m_A \rangle \simeq 73 \, {\rm GeV}$ 
this scale is $Q_E \simeq 1.5 \, (m_X / (m_X + m_A))^2 \, {\rm keV}$. Let us 
also define the function
\begin{equation}
D(q) \equiv \frac{1}{I_n (5.2)} \int_{x_{\rm min} (q)}^{5.2} dx \, \frac{\Theta 
(x-1)}{x^{0.9} \, \sqrt{x-1} \, \sqrt{3 - 2 \, x^{-1}}} \ , \ x_{\rm min} (q) 
\equiv \frac{2}{3-q} \, , \label{eq5.24}
\end{equation}
where $I_n (5.2) = 2.3474$ is the integral (\ref{eq5.10}). In terms of these 
definitions, the figure of merit (\ref{eq5.21}) reads
\begin{mathletters}
\label{eq5.25}
\begin{eqnarray}
\rho (Q) & = & \rho_1 \, D \left( \frac{Q}{Q_E} \right) \, , 
\label{eq5.25a} \\
&& \nonumber \\
\rho_1 & = & \frac{\sqrt{\pi}}{2} \, \frac{v_o}{v_E} \, \delta_E = 1.39 \, 
g_{\rm tot}^{(-10)} \, . \label{eq5.25b}
\end{eqnarray}
\end{mathletters}
The function $D(q)$, with $q \equiv Q / Q_E$ is plotted in Figure 2.

 The 
plateau that is reached by $D(q)$ as soon as $q \leq 1$ (i.e. $Q \leq
Q_E$) has 
 value $D(1) = 0.803$. This gives the maximal figure of merit \cite{damkrauss1}
\begin{equation}
\rho (Q_E) = 1.11 \, g_{\rm tot}^{(-10)} \, . \label{eq5.26}
\end{equation}
If $g_{\rm tot}^{(-10)} \sim 1$ (see below), this yields a 100~\% increase of 
the differential event rate below $Q = Q_E \sim {\rm keV}$.

Also, within our present rough 
approximation, $e \approx 1$, the direction of the incoming WIMPs from the new 
population will be {\it strongly anisotropic}. Indeed, not only are they 
entirely confined in the ecliptic plane, but even in this plane they have 
velocities whose local direction is within $\pm \, {\rm tan}^{-1} \, \sqrt{2 
(1-1/5.2)} = \pm \, 51.8^{\rm o}$ of the vector $- {\bf v}_E$. This directional 
information might greatly help in distinguishing the real events from the 
background if one had a directional detector. Note, however, that long-term
numerical simulations of the evolution  of the elliptic elements of the WIMPs
are probably needed to assess the robustness of this prediction, and that for
the WIMP spectrum, based on the crude estimate
$e \approx 1$. For instance, one expects the actual spectrum $\rho (Q)$ to be
a somewhat smoothed  version of Figure 2, though the existence of a hump
around $Q_E$ should survive.

In view of this uncertainty on the exact spectrum of the WIMP population, we did 
not compute a precise figure of merit for the indirect neutrino signal from the 
Earth. Let us only point out the main features of the new signal. First, the 
capture by the Earth of a slow population $v_X^{\rm new} \sim v_E$, instead of 
the standard $v_X^{\rm standard} \sim v_o$, is more effective. If we take only 
this effect in account, one would expect a figure of merit of order $(v_o / v_E) 
\, \delta_E \sim \rho_1 \sim 1.4 \, g_{\rm tot}^{(-10)}$.
 However, another effect 
is also quite important. Because of the large ratio $(v_o / v_{\rm escape}^{\rm 
Earth})^2 \sim (220 \, {\rm kms}^{-1} / 11 \, {\rm kms}^{-1})^2 \sim 400$ the 
Earth capture probability for incoming standard WIMPs is strongly peaked around 
the ``resonances'' $m_X \simeq m_A$ for some element $A$ in the Earth 
\cite{gould87}. For the new population $(v_X / v_{\rm escape}^{\rm Earth})^2 
\sim 9$ is much smaller, and the resonances become much broader. This means in 
particular that for masses $m_X > m_{56}$ (iron resonance) the neutrino signal 
will be much amplified, compared to standard expectations.
However, the maximum WIMP mass for which capture is important depends
very sensitively on the low-velocity cut-off $v_c$ (measured in the Earth
frame) of the WIMP population. Indeed,
 taking into account the fact that the escape velocity from the iron core
of the Earth is $ v_{\rm esc}^{\rm Fe} \sim 15 {\rm km} {\rm s}^{-1}$, 
 capture is only possible when 
$\beta_{-}^{\rm Fe} \geq (v_c/v_{\rm esc}^{\rm Fe})^2 $, which
defines an upper bound on $m_X$. For instance, within our present (rough)
approximation, $v_c = v_{\oplus}$ so that only WIMPs with mass 
$m_X \leq 2.62 m_{\rm Fe} \simeq 147 
{\rm GeV}$ would be captured by the Earth.

\section{Estimates for Realistic WIMPs}

To determine the relevance of the 
effects discussed here to the ongoing search for halo WIMPs, the actual
numerical value of
$g_{\rm  tot}^{(-10)}$ for realistic WIMPs is of central importance to consider.
To investigate this question we have explored the parameter space  of the
theoretically best motivated WIMP candidate: the lightest supersymmetric 
particle (assumed to be a neutralino) of the ``Minimal Supersymmetric Standard 
Model'' (MSSM) \cite{jungman}.  Of course this is really not a single model, but
a range of models, depending upon the assumptions one makes about such issues as
Unification, and also the nature of Supersymmetry breaking. Because of this,
detailed specific predictions of remnant neutralino densities, elastic
scattering cross sections, etc., are difficult to give with any generality. 
Indeed, our understanding of SUSY models is still developing, so that
predictions of annihilation rates in the early universe, and thus remnant
neutralino densities may require alteration \cite{kane}.  

In any case, for the purposes of this investigation it is worth exploring
the general order of magnitude of predicted solar capture cross sections, and
the resulting solar system density of SUSY WIMPs.  To this end, to sample
the many SUSY parameters, we have made use of the
specialized code Neutdriver 
written by Jungman, Kamionkowski and Griest
\cite{jungman}.  This allows a calculation, using a specific parameterization
of MSSMs, of annihilation rates, remnant neutralino densities, and elastic
scattering cross sections on isotopes in the Sun, and in potential terrestrial
detectors.  

Constraints on the SUSY parameter space are also model dependent, depending on
whether one uses various GUT relations for gaugino masses, and also on
assumptions about universality of scalar and fermion masses.  These constraints
are also evolving as new data from LEP, and from such processes
as $ b \rightarrow s \ \gamma $ are obtained.  At the time the calculations
reported here were performed, these constraints led us to sample the SUSY phase
space described in Table 1 (conventions are those of Jungman ${\it et al.}$
\cite{jungman}).

As implied by the data in Table 1, a total of 9600 different sets of SUSY
parameters were initially chosen.  Among these some
combinations, reported by Neutdriver, were unphysical or phenomenologically
unacceptable for a variety of reasons.  These were culled, and the remaining
allowed configurations were utilized to determine remnant densities and
scattering cross sections. 

This residual phase space has some characteristics that are important to
distinguish here.  First, we include neutralino masses as large as 400 GeV. 
These masses are larger than conventionally displayed in constraints by ongoing
direct WIMP detection experiments.  However as we are interested here primarily
in knowing how broadly relevant our results might be, we wanted to explore as
large a region of phase space as possible ---independent of model builders' or
experimentalists' preferences.  This factor also played a role in our choice of
WIMP cosmic densities to include in this analysis.  From the broad phase space
that survived the above cuts, we then selected those models that resulted in
a remnant density in the range $0.025< \Omega_Xh^2 < 1$.  This range again is
somewhat broader than is conventionally chosen for $\Omega$.  However, given
the upper limit $\Omega_{\rm Baryon}h^2 < .026$ from Big Bang
Nucleosynthesis \cite{krausskern} it 
remains
 possible that the non-baryonic abundance could be
as small as the lower bound quoted above and still at least marginally exceed
the baryon density in our own galaxy.   This relaxed choice of density
restriction, combined with the higher mass range we consider, implies that we
allow models with somewhat higher cross sections on terrestrial targets than is
usually displayed when SUSY constraint diagrams are displayed.  (It is
generally true, for example, that those models with the lowest remnant density
today have the highest elastic scattering cross sections, for reasons made
clear in the introduction to this paper.). In any case, the results quoted here
are meant to be indicative of what one might expect for realistic WIMPs, and
since SUSY model predictions are themselves evolving, the detailed model
results quoted here should be taken as indicative of the general order of
magnitude of one's expectations for SUSY WIMPs.  Nevertheless, in order to
explore how restricting the remnant neutralino density will affect the range of
$g$'s expected for SUSY WIMPs, we also considered subset of the parameter space 
in
which $0.1<
\Omega_Xh^2 < 1$.

Each neutralino has two possible modes of scattering on targets, in both the
Sun and in terrestrial detectors.  Because neutralinos are Majorana particles
the non-relativistic limit of the scattering cross section generically involves
a spin-dependent piece.  In addition, exchange of scalar particles can produce
a scalar, spin-independent piece.  This latter term, if present, generally
dominates for large nuclear targets, because in this case the WIMP can scatter
coherently off of the entire nucleus with a coupling proportional to the atomic
number of the nucleus $A$.  Thus the cross section goes as $A^2$.  Moreover,
the cross section generically involves as factor the square of the WIMP-nucleus
reduced mass, $(m_A m_X/(m_A + m_X))^2$, which, for
heavy WIMPs, also increases as the square of the mass of the
target nuclei.  Since this latter quantity is also proportional to $A$, this
implies that the scattering cross section for heavy nuclei can include a factor
proportional to $A^4$, which for nuclei as heavy as iron or germanium can be
very large indeed.   

As a result, one generically finds that scattering cross sections are dominated
by the coherent scalar piece in all cases except where this piece is suppressed
due to various model-dependent factors. Moreover, with the exception of hydrogen
 and nitrogen, all other nuclei in the
Sun are even-even nuclei, and for these nuclei the spin-dependent amplitude
vanishes. In calculating the solar capture rate
described in this paper, we included all known elements in the Sun, with
abundances given by current solar model calculations. All elemental abundances
except for hydrogen and helium are taken from Jungman ${\it et al.}$
\cite{jungman} while the former two abundances are taken from Bahcall and
Pinsonneault \cite{bahcall}.  We display the mass fractions used for the
various elements in Table 2.

In addition to calculating the capture factors $g_A$, we also determined the
scattering cross sections on germanium, which is currently the target material
of choice in cryogenic detectors.  In order to express the results in a more
target independent way, however, we adopt a standard presentation of this cross
section in terms of the effective WIMP-nucleon cross section. This is obtained
by scaling from targets of atomic number $A$, using the assumption of coherent
scattering, and is given by:

\begin{equation}
\sigma_p \equiv {\sigma_A \over A^2} ({m_X + m_A \over m_X m_A})^2
({m_X m_p \over m_X + m_p})^2 \simeq
 {\sigma_A \over A^4} {(m_X + m_A)^2 \over (m_X + m_p)^2} \, .
\end{equation}

We present our results in Tables  3 and 4 and Figures 3-6.  In the tables, in
addition to  the value of $g_{\rm tot}^{(-10)}$, we also display the value of
$g_{\rm H_{\rm scalar}}^{(-10)}$, $g_{\rm Fe}^{(-10)}$, 
and $g_{\rm O}^{(-10)}$.  In the
figures we display WIMP-Nucleon effective cross sections as a function of mass
for different models, where models with
$\mu >0$ ($\mu <0$) are displayed in odd (even) figures.  The size of the model
point gives the range of the value of
$g_{\rm tot}^{(-10)}$ calculated for this model. The different figures refer to
different cutoff values for the WIMP remnant cosmic density. In the figures we
also show approximate limits obtained from direct detection experiments on
$\sigma_p$
\cite{cdms} under two assumed values for the local halo WIMP density
($\rho = 0.3 {\rm GeV cm^{-3}}$ and $\rho = 0.1 {\rm GeV cm^{-3}}$). 

Several features should be clear from these results.  
First, $g_{\rm tot}$ values in
excess of unity are clearly possible, implying that realistic WIMPs to which
the next generation of direct detectors will be sensitive should be expected to
have a solar-system density in the region of the Earth that is significant.
Second, we note that there is in general a
monotonic relation between $\sigma_p$ and $g_{\rm tot}$,
 with however, wide dispersion.
Approximately one has

\begin{equation}
 g_{\rm tot}^{(-10)} \sim \sigma_p/(6 \times 10^{-41} {\rm
cm^2}) \, . 
\end{equation}

 Also note that in this case, the dominant single contribution to
$g_{\rm tot}$ comes from scattering on iron
 in the Sun, although the net contribution
to $g_{\rm tot}$ from the combination of lighter 
elements is of the order of $ 40\%$ of the total.  For $\mu <0$, an additional
possibility arises, at least for the lowest values of
$\Omega_Xh^2$ and for low mass WIMPs.  In this case, solar capture can be
dominated by spin-dependent scattering off  hydrogen so that the
germanium cross section, and hence the effective
$\sigma_p$ can be several orders of magnitude smaller than in the case of
WIMPs for which the dominant scattering on heavy nuclei is coherent.

As one increases the lower limit on the remnant WIMP cosmic density today,
several effects ensue.  First, as expected, increasing this lower cutoff tends
to decrease the mean value of $g_{\rm tot}$.   Also, for $\mu <0$
the low mass low
$\sigma_p$ branch of WIMP phase space rapidly decreases in size, disappearing
completely by the time the cutoff on $\Omega{h^2}$ exceeds 0.1.  Thus, if the
cosmic density exceeds this value, then a large solar system population is in
one to one correspondence with WIMPs that should be directly detectable in the
next generation of detectors.

Finally, we have examined what would be the effect of changing the 
average velocity dispersion of halo WIMPs.  We considered a range $180 < v_o <
270$, which encompasses most estimates for this quantity for our galactic
halo, and found that the results in all cases changed by less than
$10\%$ compared to those quoted above.  It is interesting that while the change
was too slight to be significant, the direction of the change was not
monotonic, but depended upon the mass of the WIMP and the dominant target atom
in the Sun.  For heavy WIMPs whose dominant scattering was on heavy elements,
increasing $v_o$ increased the capture factor $g_{\rm tot}$.

\section{Conclusions}

The results presented here are quite encouraging, and motivate a consideration
of detection schemes that might probe down to keV energy deposits by WIMP
scattering.  If this is possible, the observation of a rise in the differential
event rate for low energy events, of the form we describe here, could provide a
very useful discriminant that could demonstrate that any claimed WIMP signal
at higher energy is, in fact, due to halo WIMPs.  While it is challenging to
consider obtaining sensitivity to such low energy events (and more importantly
reducing the background of noise for such events), this may be less daunting
than attempting to achieve directional sensitivity, which is the alternative
discriminant that has been discussed \cite{spergel,copikrauss}.  Of course, if
one had directional sensitivity, the signal we discuss here should be even
easier to disentangle from backgrounds, as we expect it should be extremely
anisotropic, as described earlier. 

Nevertheless, the analytical results presented here are in some sense still
preliminary.  While we expect the general quantitative features of this new
WIMP population will be well approximated by the results presented here,
full scale numerical simulations of the WIMP orbits under consideration will be
necessary to confirm the details of our results.  In particular, knowledge of
the anisotropy of the distribution, as well as its energy spectrum will require
such simulations, and the results presented here should be considered
qualitative in these regards.  In addition, such simulations, which incorporate
the presence of the planets and allow close encounters, will be necessary to
confirm that the WIMP population we focus on here is indeed long-lived in the
solar system, and that its spatial distribution is not 
critically different
from the simple spherically symmetric, high-eccentricity one we have assumed.

One area that has not been investigated in detail here, and which certainly
warrants further investigation, is the implications of this new distribution
for indirect WIMP detection via annihilations in the Earth.  As we described in
the text, it is quite likely that this signal could be significantly enhanced,
especially for heavy WIMPs, compared to that which is calculated for halo WIMP
capture by the Earth.  We expect, in fact, that new bounds on SUSY phase space
may be possible on the basis of such considerations, compared to existing
limits from underground neutrino detectors.  Such an investigation is currently
underway.

\acknowledgments 
We thank G.~Jungman for kindly providing us with a copy of the code
Neutdriver. The hospitality of CERN, where this work was initiated,
is gratefully acknowledged. LMK thanks the colleagues at CERN 
for discussion on SUSY limits from LEP.  In
addition, he thanks the IHES for hospitality during the completion of this
project. We thank Andrew Gould for suggesting a simplified derivation
of Eq.~(\ref{eq2.20}).
 We also wish to thank D. Devaty and P. Kernan for
significant programming support and for discussions,
and the anonymous referee for constructive and insightful comments. 
 LMK's research is
supported in part by the DOE.

\vskip 0.1in
\hskip 1.0in
\begin{table}
\begin{tabular}{||l|lr||}  \hline
$\mu < 0$  & $M_2$:  80-800 GeV &(10 steps) \\
& $M_1,M_3$  determined by GUT relations & \\
& ${\mu}$: -800 - -60 GeV &(10 steps)  \\
& ${\rm tan} \beta$:  2 -40 &(4 steps) \\
& $m_A$ :  70 -500 GeV &(3 steps) \\
& $m^2_{\rm squark}$:  $4 - 64 \times 10^4 \, {\rm GeV}^2$ &(4 steps) \\ \hline
$\mu > 0$  & $M_2$:  80-800 GeV &(10 steps) \\
& $M_1,M_3$  determined by GUT relations & \\
& ${\mu}$: 150 - 800 GeV &(10 steps)  \\
& ${\rm tan} \beta$:  2 -40 &(4 steps) \\
& $m_A$ :  70 -500 GeV &(3 steps) \\
& $m^2_{\rm squark}$:  $4 - 64 \times 10^4 \, {\rm GeV}^2 $ &(4 steps) \\ \hline
\end{tabular}
\caption{SUSY Parameter Space Sampled in Estimates}
\end{table}

\vskip 0.1in
\hskip 2.0in
\begin{table}
\begin{tabular}{||l||c||l||}  \hline
Element & Atomic Mass Number & Mass Fraction \\ \hline
H & 1 & 0.7095 \\
He & 4 & 0.2715 \\
C & 12 & 3.87 $\times 10^{-3}$ \\
N & 14 & 9.40 $\times 10^{-4}$ \\
O & 16 & 8.55 $\times 10^{-3}$ \\
Ne & 20 &1.51 $\times 10^{-3}$ \\
Mg & 24 & 7.39 $\times 10^{-4}$ \\
Si & 28 & 8.13 $\times 10^{-4}$ \\
S & 32 & 4.65 $\times 10^{-4}$ \\
Fe & 56 & 1.46 $\times 10^{-3}$ \\ \hline
\end{tabular}
\caption{Elemental Mass Fractions Used in Solar Capture Estimates}
\end{table}

\vskip 0.1in
\begin{table}
\begin{tabular}{||c||c|c|c|c|c|c|c||}  \hline
& $\Omega{h^2}$ &   $m_{\rm X}$(GeV) & $g_{\rm tot}$ & $g_{\rm H}$ & 
$g_{\rm O}$ & $g_{\rm Fe}$ & $\sigma_p$
  \\ \hline
$\mu > 0$ & & & & & & & \\ \hline
& 0.037 & 384 & 2.480 & 0.008 & .410 & 1.532 & 1.38E-40 \\ 
& 0.038 & 384 & 2.239 & 0.007 & .370 & 1.381 & 1.25E-40 \\
& 0.036 & 384 & 2.219 & 0.007 & .367 & 1.370 & 1.24E-40 \\
& 0.032 & 306 & 1.943 & 0.008 & .354 & 1.136 & 1.27E-40 \\   
& 0.029 & 306 & 1.886 & 0.007 & .342 & 1.103 & 1.24E-40 \\ 
& 0.027 & 306 & 1.871 & 0.007 & .340 & 1.094 & 1.22E-40 \\ 
& 0.076 & 353 & 1.230 & 0.004 & .210 & .746 & 7.28E-41 \\  
& 0.074 & 353 & 1.173 & 0.004 & .201 & .710 & 6.94E-41 \\ \hline
$\mu < 0$ & & & & & & & \\ \hline
& 0.039 & 357 & 0.63 & 0.002 & 0.11 & 0.38 & 3.70E-41 \\ 
& 0.039 & 357 & 0.61 & 0.002 & 0.10 & 0.37 & 3.55E-41 \\
& 0.038 & 357 & 0.60 & 0.002 & 0.10 & 0.37 & 3.53E-41 \\
& 0.025 & 32 & 0.59 & 0.000 & 0.00 & 0.00 & 8.17E-44 \\
& 0.037 & 276 & 0.59 & 0.002 & 0.11 & 0.34 & 4.16E-41 \\
& 0.035 & 276 & 0.58 & 0.002 & 0.11 & 0.33 & 4.08E-41 \\
& 0.034 & 276 & 0.58 & 0.002 & 0.11 & 0.33 & 4.08E-41 \\
& 0.029 & 196 & 0.55 & 0.003 & 0.12 & 0.29 & 5.12E-41 \\ \hline
\end{tabular}
\caption{Largest $g_{\rm tot}^{(-10)}$ values for $0.1 > \Omega_X h^2 > 0.025$}
\end{table}

\vskip 0.1in
\begin{table}
\begin{tabular}{||c||c|c|c|c|c|c|c||}  \hline
& $\Omega{h^2}$ &   $m_{\rm X}$(GeV) & $g_{\rm tot}$ & $g_{\rm H}$ & 
$g_{\rm O}$ & $g_{\rm Fe}$ & $\sigma_p$
  \\ \hline
$\mu > 0$ & & & & & & & \\ \hline
& 0.135 & 397 & 1.11 & 0.003 & 0.18 & 0.69 & 6.06E-41 \\
& 0.150 & 80 & 0.97 & 0.013 & 0.23 & 0.47 & 2.07E-40  \\
& 0.126 & 80 & 0.96 & 0.013 & 0.23 & 0.47 & 2.06E-40  \\
& 0.211 & 80 & 0.86 & 0.012 & 0.20 & 0.42 & 1.84E-40  \\
& 0.143 & 397 & 0.74 & 0.002 & 0.12 & 0.46 & 4.07E-41 \\
& 0.136 & 397 & 0.74 & 0.002 & 0.12 & 0.46 & 4.04E-41 \\
& 0.110 & 316 & 0.64 & 0.002 & 0.12 & 0.38 & 4.10E-41 \\
& 0.103 & 316 & 0.62 & 0.002 & 0.11 & 0.37 & 3.97E-41 \\
& 0.165 & 359 & 0.46 & 0.002 & 0.08 & 0.28 & 2.69E-41 \\
& 0.170 & 359 & 0.43 & 0.001 & 0.07 & 0.26 & 2.54E-41 \\
& 0.164 & 359 & 0.43 & 0.001 & 0.07 & 0.26 & 2.50E-41 \\
& 0.123 & 396 & 0.39 & 0.001 & 0.06 & 0.24 & 2.14E-41 \\
& 0.124 & 278 & 0.37 & 0.002 & 0.07 & 0.21 & 2.63E-41 \\ \hline
$\mu < 0$ & & & & & & & \\ \hline
& 0.101 & 361 & 0.26 & 0.001 & 0.04 & 0.16 & 1.49E-41 \\ 
& 0.104 & 361 & 0.24 & 0.001 & 0.04 & 0.15 & 1.40E-41 \\ 
& 0.107 & 199 & 0.21 & 0.001 & 0.04 & 0.11 & 1.92E-41 \\ 
& 0.148 & 402 & 0.18 & 0.001 & 0.03 & 0.11 & 9.61E-42 \\ 
& 0.149 & 402 & 0.18 & 0.001 & 0.03 & 0.11 & 9.53E-42 \\ 
& 0.130 & 321 & 0.18 & 0.001 & 0.03 & 0.10 & 1.10E-41 \\ 
& 0.135 & 321 & 0.17 & 0.001 & 0.03 & 0.10 & 1.06E-41 \\ 
& 0.134 & 321 & 0.17 & 0.001 & 0.03 & 0.10 & 1.06E-41 \\ 
& 0.168 & 362 & 0.16 & 0.001 & 0.03 & 0.10 & 9.17E-42 \\ 
& 0.169 & 240 & 0.16 & 0.001 & 0.03 & 0.09 & 1.24E-41 \\ \hline
\end{tabular}
\caption{Largest $g_{\rm tot}^{(-10)}$ values for $1.0 > \Omega_X h^2 > 0.1$}
\end{table}

\begin{figure}[htb]
\begin{center}\
\epsfxsize = 5.0in \epsfbox{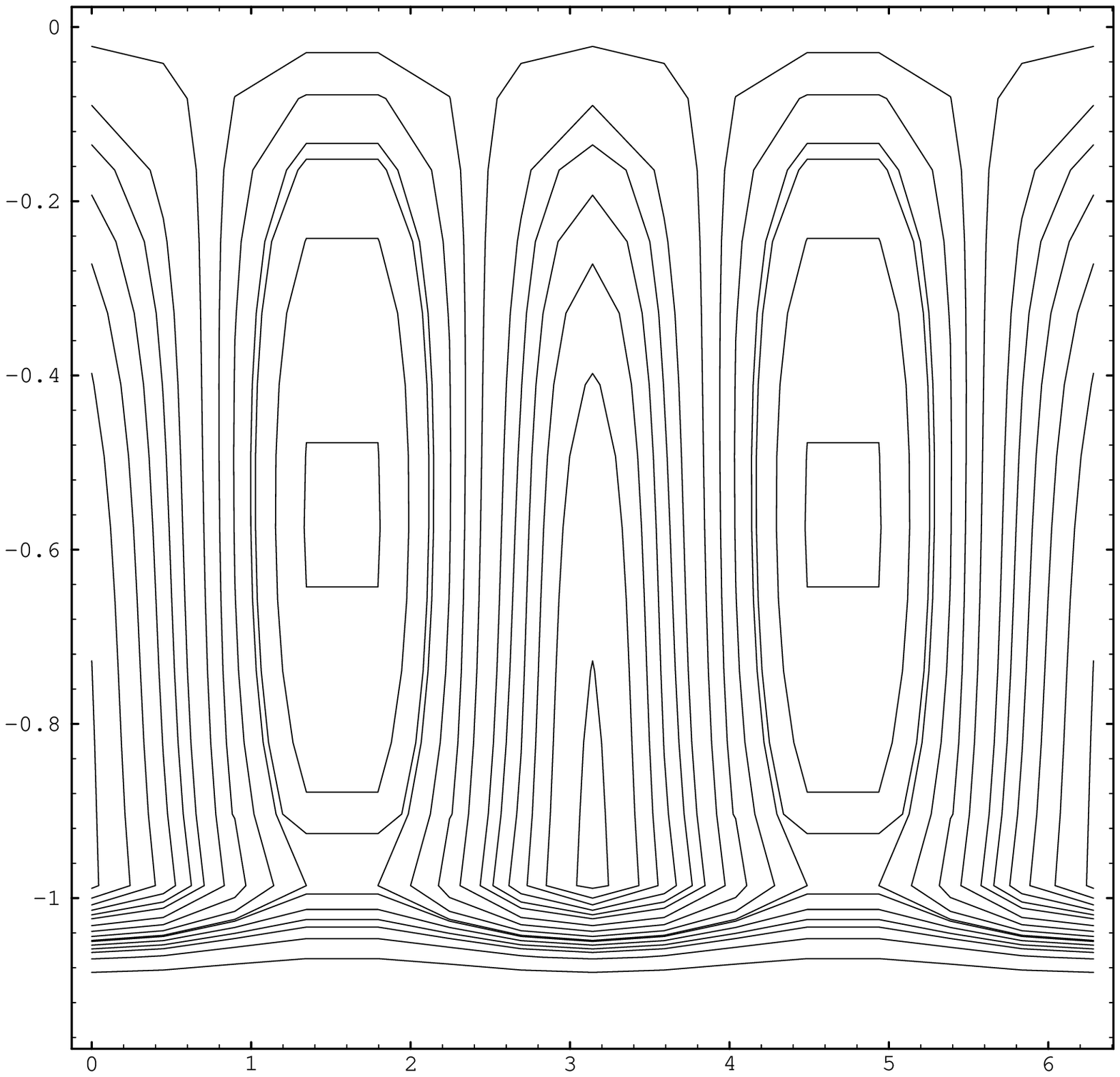}
\end{center}
\vskip -4.0in
{\hskip 0.5in \rotate[l]{$log_{10} \sqrt{1-e^2}$}}
\vskip 1.75in
\begin{center}
{\hskip 0.5in {$g$ (in radians)}}
\end{center}
\caption{Level curves of the perturbation Hamiltonian describing the secular
evolution of the canonical pair $(G,g)$. 
 Note the divide between trajectories
that always stay within the Sun  and
those that get out.}
\end{figure}

\begin{figure}[htb]
\begin{center}\
\epsfxsize = 5.0in \epsfbox{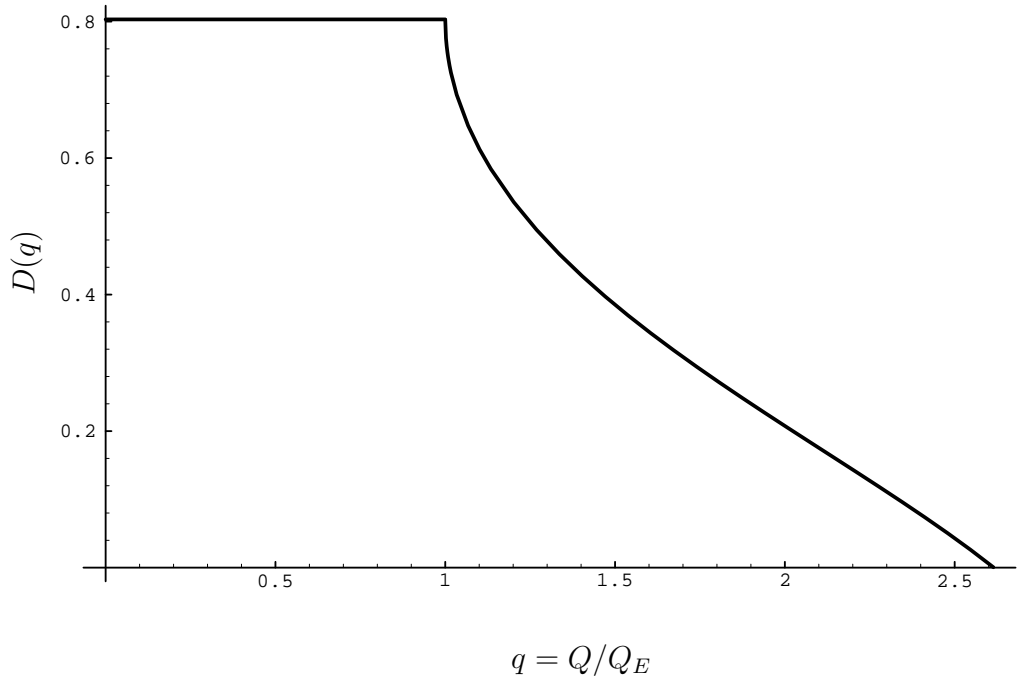}
\end{center}
\vskip -4.0in
{\hskip 0.5in \rotate[l]{$D(q)$}}
\vskip 1.75in
\begin{center}
{\hskip 0.5in {$q=Q/Q_E$ }}
\end{center}
\caption{ Shape of the differential rate ( per keV) of the additional scattering
events caused by the  WIMP population considered here. This figure represents
the function $D(q)$, where $q = Q/Q_E$ is the energy transfer in units of the
characteristic energy scale $Q_E$ (in the  keV range).}
\end{figure}

\begin{figure}[htb]
\begin{minipage}[t]{77mm}
\epsfxsize = \hsize \epsfbox{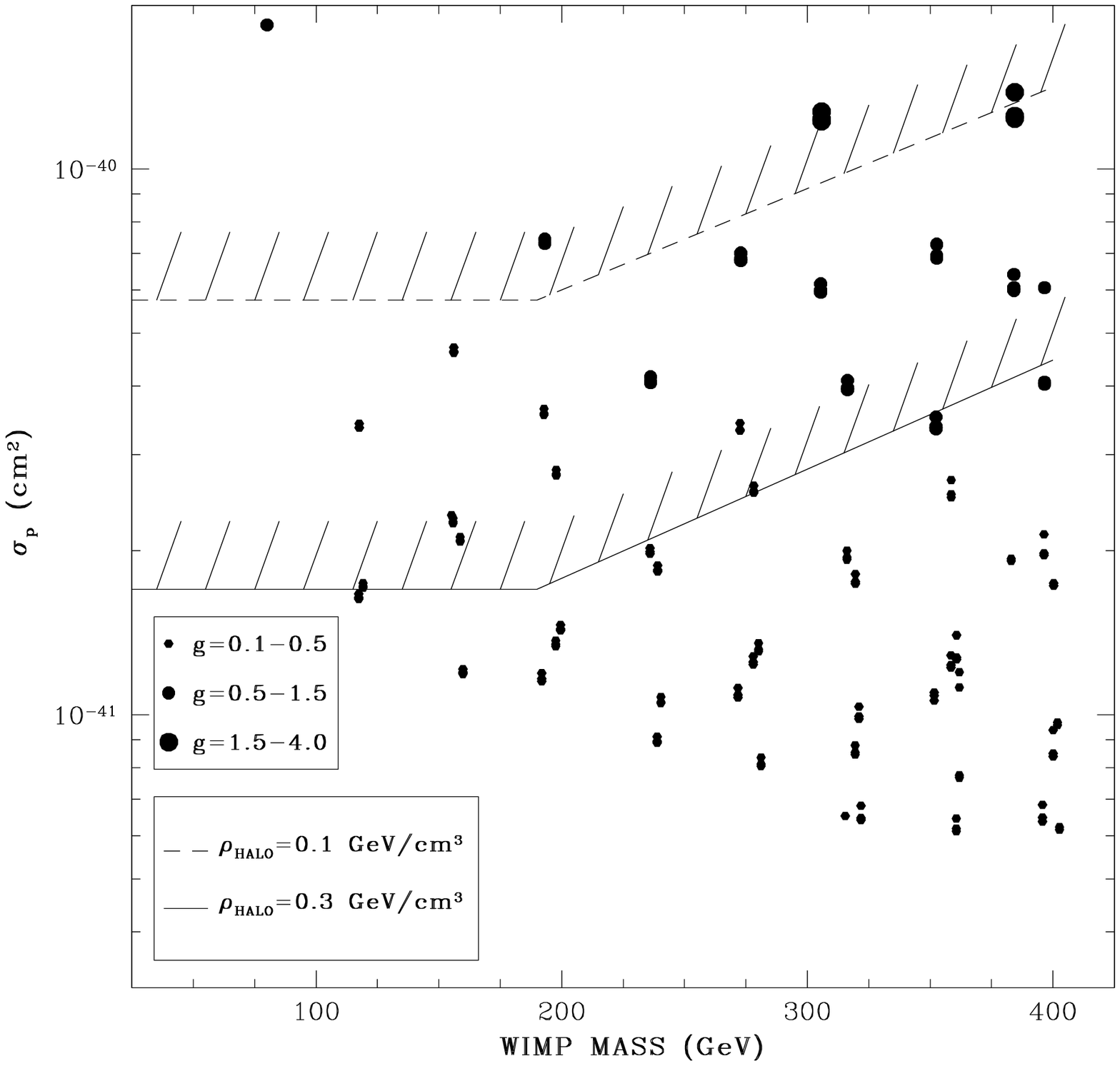}
\caption{$g_{\rm tot}$ as a function of the effective
WIMP nucleon cross section, $\sigma_p$, and WIMP Mass,
for
$\mu > 0$, and assuming $\Omega_X h^2 > 0.025$.  Hatched curves represent
experimental upper limits assuming $\rho = 0.3 {\rm GeVcm^{-3}}$ (lower)
and $\rho = 0.1 {\rm GeVcm^{-3}}$ (upper).}
\end{minipage}
\hspace{\fill}
\begin{minipage}[t]{77mm}
\epsfxsize = \hsize \epsfbox{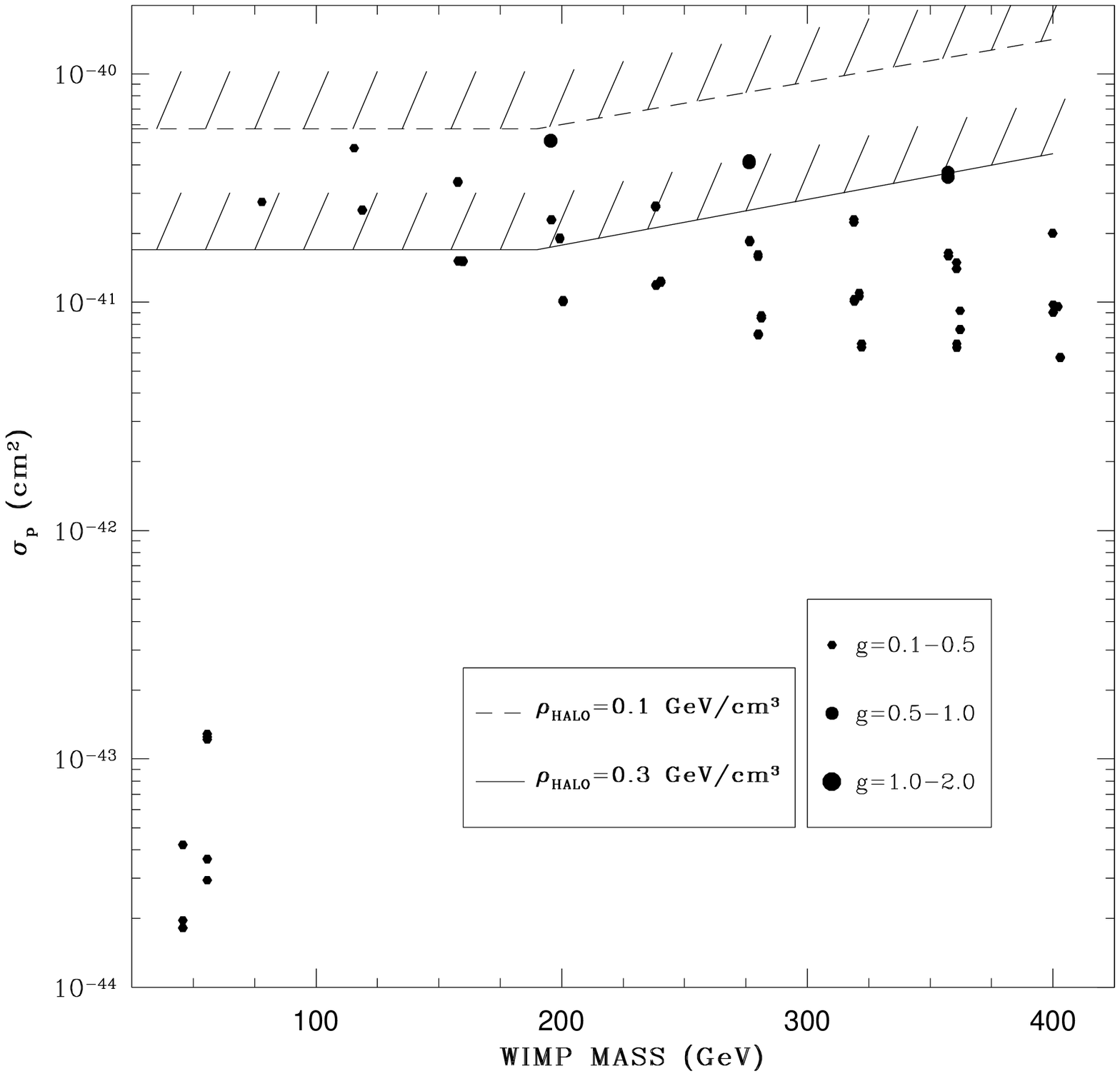}
\caption{Same as Figure 3 but with $\mu < 0$}
\end{minipage}
\end{figure}

\begin{figure}[htb]
\begin{minipage}[t]{77mm}
\epsfxsize = \hsize \epsfbox{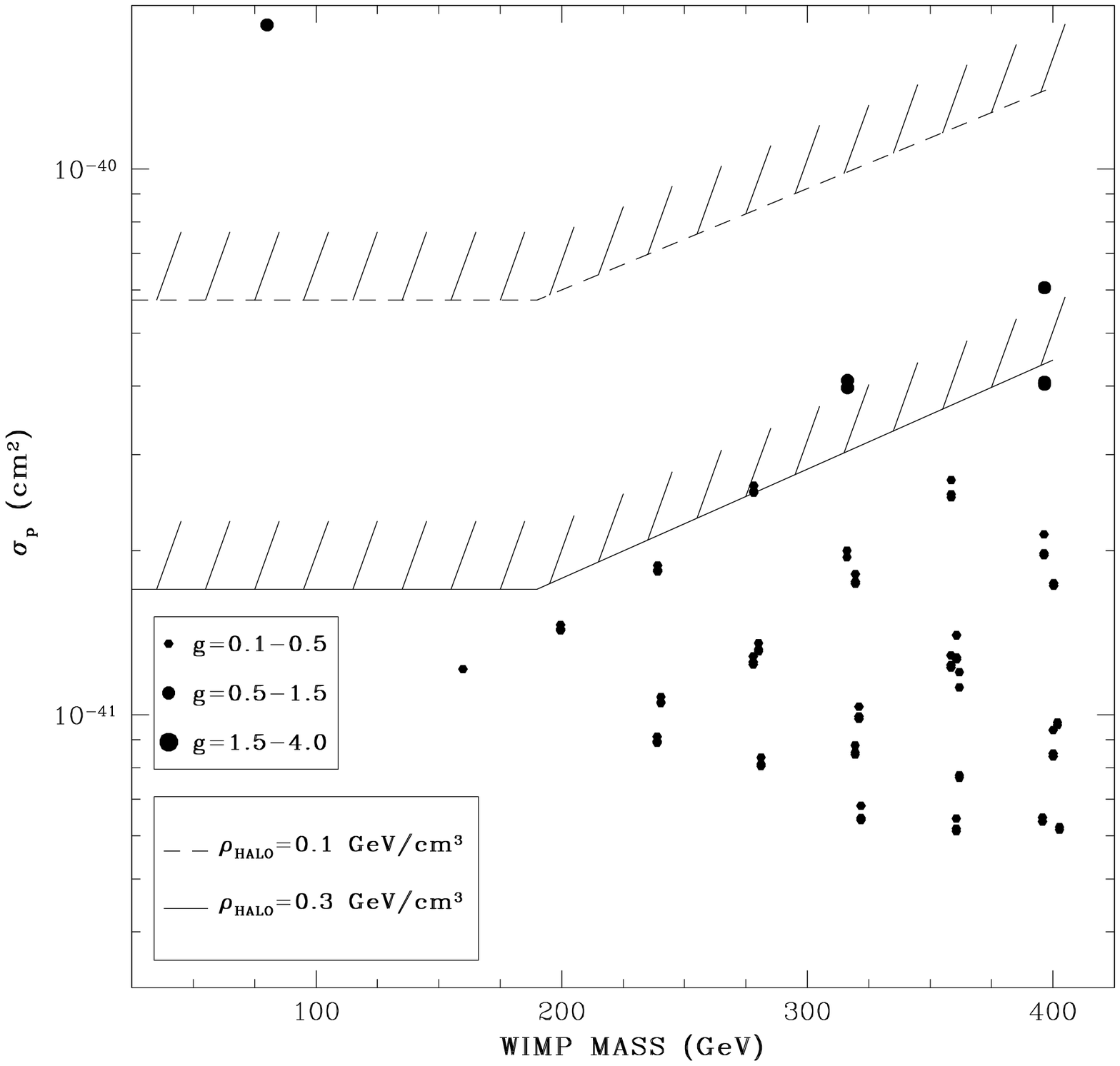}
\caption{Same as Figure 3, but assuming $\Omega_X h^2 > 0.1$}
\end{minipage}
\hspace{\fill}
\begin{minipage}[t]{77mm}
\epsfxsize = \hsize \epsfbox{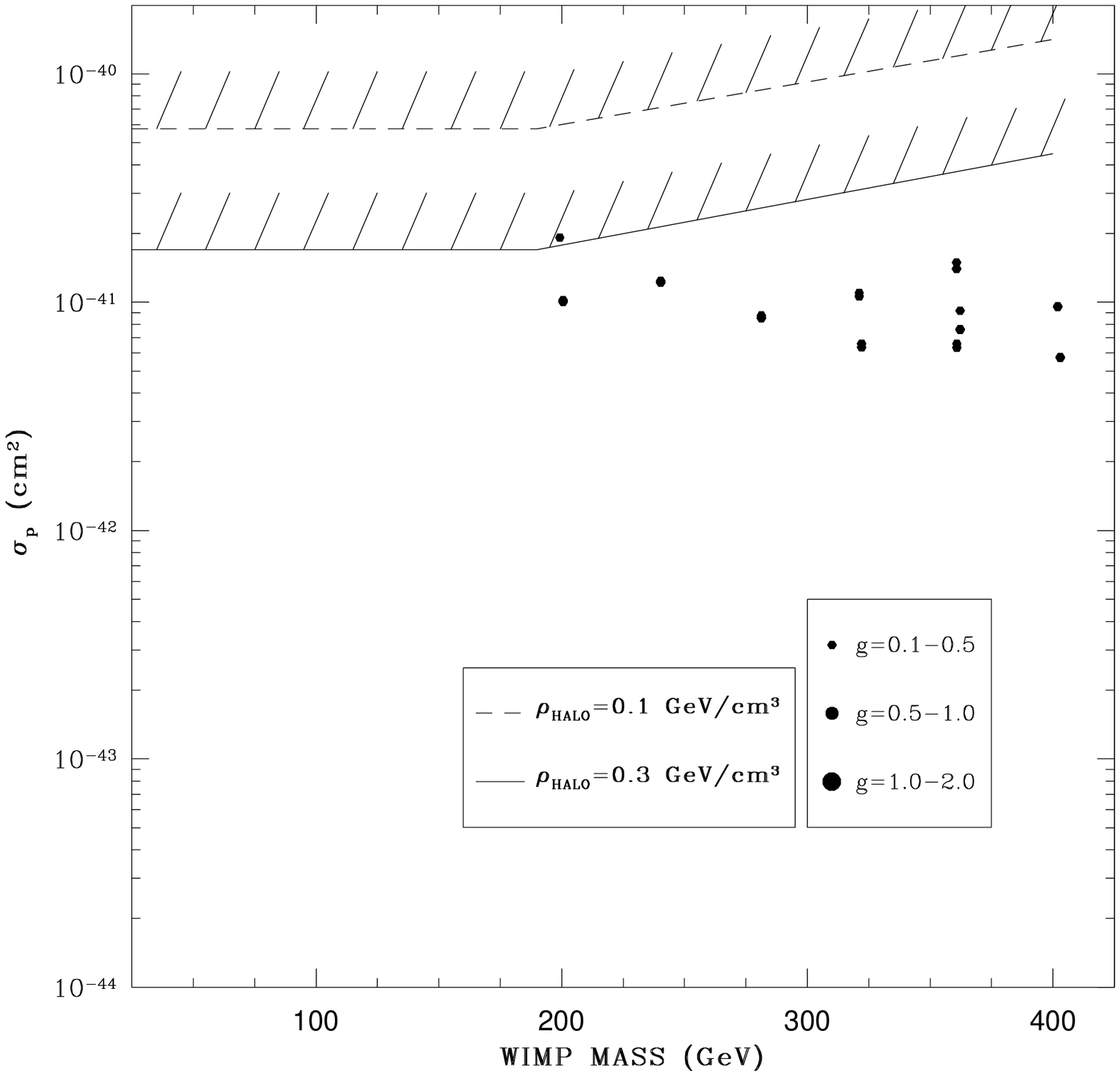}
\caption{Same as Figure 5, but with
$\mu < 0$}
\end{minipage}
\end{figure}

\end{document}